\documentstyle[11pt,epsfig]{article}
\begin{document}
\setlength{\baselineskip}{0.30in}
\newcommand{\beq}{\begin{equation}}
\newcommand{\eeq}{\end{equation}}
\newcommand{\be}{\begin{eqnarray}}
\newcommand{\ee}{\end{eqnarray}}
\newcommand{\rar}{\rightarrow}
\newcommand{\lar}{\leftarrow}
\newcommand{\lrar}{\leftrightarrow}
\newcommand{\bi}{\bibitem}
\newcommand{\cd}{\cdot}
\newcommand{\vp}{\varphi}
\newcommand{\vpc}{\varphi_c}
\newcommand{\tvpc}{\tilde \varphi_c}

{\hbox to\hsize{October, 1998 \hfill TAC-1998-029}
\begin{center}
\vglue .06in
{\Large \bf {Equation of motion of a classical scalar field with back reaction
of produced particles. }}
\bigskip
\\{\bf A.D. Dolgov
\footnote{Also: ITEP, Bol. Cheremushkinskaya 25, Moscow 113259, Russia.}
\footnote{e-mail: {\tt dolgov@tac.dk}}, 
S.H. Hansen\footnote{e-mail: {\tt sthansen@tac.dk}}
 \\[.05in]
{\it{Teoretisk Astrofysik Center\\
 Juliane Maries Vej 30, DK-2100, Copenhagen, Denmark
}}}
\\[1.0in]

\end{center}
\begin{abstract}

In the one-loop approximation we derive the equation of motion for a 
classical scalar field $\vpc (t)$ with the back reaction of particle 
production included. 
Renormalization of mass and couplings of $\vpc$ is done  explicitly. The 
equation is non-local in time, but can easily be treated perturbatively or 
numerically. For the weak trilinear coupling of the external field to the 
produced particles, the new equation gives the same solution as the familiar 
one with the $\Gamma \dot \vpc$ term. For a stronger coupling and other types 
of couplings the results are significantly different. The equation can be 
applied to the universe heating by the inflaton decay and to spontaneous 
baryogenesis. 
 
\end{abstract}

PACS: 98.80.Cq, 14.80.-j, 95.30.Cq

\section{Introduction}

In the classical theory of particle production  (see  
e.g.~\cite{gmm}) the external time-dependent field is assumed to be "fixed", 
i.e. not affected by the back reaction of the produced particles. Though it is
a very good approximation for the situation that can be realized under
laboratory conditions, it is not so in cosmology, when an external cosmic
field decays and transfers energy to the produced particles.
Usually the  back reaction is described by an addition of 
the term $\Gamma \dot \vpc$
into the usual Klein-Gordon equation governing the evolution of a 
(spatially homogeneous, though not necessarily so) classical scalar
field:
\beq
\ddot{\vp_c} + 3 H \dot{\vp_c} + V'(\vp_c) =  -\Gamma \dot {\vp_c}~,
\label{eqgam}
\eeq
where $H=\dot a /a$ is the Hubble parameter describing the expansion of the 
universe (this term is absent in flat space-time), $V(\vpc)$ is the potential
of the field $\vpc$, the prime denotes the derivative with respect to
the field $\vpc$, and $\Gamma >0$ is the decay width. One argument in
favour of a modeling  the back reaction in this way is that 
the solution of eq. (\ref{eqgam}), 
in the case of the simplest harmonic potential, $V(\vpc) = m^2 \vpc ^2 /2$,
and for the trilinear coupling to the produced particles,
${\cal L}_{int} \sim \vpc \chi^* \chi$, has the form:
\be
\vp_c (t) = \vp_0 \exp (-\Gamma t /2) 
\cos \left(\sqrt{m^2 -\Gamma^2/4}\,\, t \right)\, ,
\label{solgam}
\ee 
which correctly describes the behaviour of $\vpc$ known from perturbation 
theory (it is assumed here and in what follows that the space-time is flat).
However, it is easy to see that already in the case of  
more complicated potentials and couplings, e.g. for
$V(\vpc) = \lambda_4 \vpc ^4 /4$ and/or ${\cal L}_{int} \sim \vpc^2 
\chi^* \chi$,
this ansatz is not applicable. Moreover,
different equations may have the same solutions and the coincidence of the
solutions in special cases does not imply the equivalence of the
theories, as we will see below in some examples. 

In what follows we will derive the equation of motion for
the classical homogeneous scalar field, $\vpc (t)$, for different 
kinds of interactions
with the produced quantum particles and different forms 
of the potential, $V(\vp)$. In fact, the back reaction term does not 
depend on the form of $V(\vp)$, but is determined only by the form 
of the interaction of $\vpc$ with the produced particles. On the other
hand, the particle production rate is sensitive to the form of the potential
$V(\vp)$.
The calculations here are made in flat space-time, but it is
straightforward to generalize them~\cite{df} to the case of the 
Friedman-Robertson-Walker metric, because the latter is conformally 
flat. The results can be applied to the universe
heating by the decay of the inflaton field and to spontaneous baryogenesis.
Our approach here is very close to that of ref.~\cite{df}.

\section{Trilinear scalar coupling}
\label{tril}

Let us first consider the simplest case of the trilinear coupling of a 
scalar field $\vp$ (which below will be taken as a classical spatially 
homogeneous field, $\vpc (t)$) to the massless scalar quantum field $\chi$. 
The Lagrangian of the fields $\vp$ and  $\chi$ has the form:
\be
{\cal L}  (\vp, \chi)=\frac{1}{2} (\partial \vp )^2 - V(\vp)
+ \frac{1}{2} (\partial \chi )^2  + f \vp \chi^2~.
\label{lphichi}
\ee
The corresponding exact quantum operator equations of motion can be written as:
\be
\ddot{\vp} -\Delta \vp + V^{'}(\vp) = f \chi^2 ,
\label{eqvp}\\
\partial^2 \chi = 2 f \vp \chi.
\label{eqchi}
\ee

We will assume that there exists a classical homogeneous field 
$\vpc (t) = \langle \vp \rangle$, where the brackets mean averaging over
quantum vacuum state of operators $\vp$ and $\chi$. The equation of motion
for $\vpc$ can be obtained by the quantum averaging of eq. (\ref{eqvp}) in 
the presence of the classical field $\vpc$~\cite{df}:
\be
\langle \ddot{\vp} + V^{'}(\vp)\rangle = \langle  f \chi^2 \rangle~.
\label{veveqvp}
\ee
Taking average of the l.h.s. is trivial, it reduces to the substitution 
$\langle \vp \rangle =\vp_c (t)$,
and averaging of the r.h.s. can be
done using the formal solution of eq. (\ref{eqchi}):
\beq
\chi(x) = \chi _0 (x) + 2 f \int d^4 y\, G^R(x, y) \, \vpc (y) \chi (y)~,
\label{soleqchi}
\eeq
where $G^R(x, y)$ is the retarded Green's function and $\chi_0$ is the 
free field operator. The latter is canonically quantized according to:
\beq
\chi (x) = \int \frac{d^3 k}{(2 \pi)^{3/2} \sqrt{2 k_0}} \left(
a_{\vec {k}}\, e^{- i k x} + a_{\vec {k}}^{\dag} \,e^{ i k x}\right)~,
\label{quant}
\eeq
where $a_{\vec{k}}$ and $a_{\vec {k}}^{\dag}$ are annihilation and 
creation operators of  momentum $\vec k$, obeying the commutation relations
$\left[ a_{\vec k} , a_{\vec k'}^\dag \right]
= (2 \pi)^3\, 2 k_0 \,\delta^3 (\vec {k} - \vec {k'})$.

The contribution of expression (\ref{soleqchi}) into the r.h.s. of
eq. (\ref{veveqvp}) gives zero result to the first order in $f$. To be 
more precise, the condition of vanishing of $\langle \chi \chi \rangle$ 
to the first order
in $f$ corresponds to elimination of tadpole diagrams and is achieved by 
a shift of $\vpc$. To the second order in $f$ the contribution is 
given by:
\be
f\langle \chi \chi \rangle = 
4 f^2 \int \frac{d^3 k}{(2 \pi)^3 2 k_0} \, \int d^4 y \,   e^{-ik(x-y)}   
  G^R (x,y) \,\vpc (y)~.
\label{firstinteg}
\ee
Using the retarded Greens function for massless scalars:
\be
G_s^R(r) = \frac{1}{4 \pi r} \, 
\delta( r -t)~,
\label{gret}
\ee
we can make almost all the integrations in eq.~(\ref{firstinteg}) and  
obtain the equation of motion of the classical scalar field, $\vpc$, with the
r.h.s. which describes the back reaction of the produced
quanta of scalar field, $\chi$, on the  evolution of the field $\vpc$:
\be
 \ddot{\vpc} + V'(\vpc)   = 
\frac{f^2}{4 \pi^2} \int_0^{t-t_{in}} \frac{d \tau}{\tau} \, \vpc(t-\tau)~,
\label{scalres}
\ee
where $t_{in}$ is an initial time, when the particle production was switched 
on (it is assumed that $t>t_{in}$).
The result is valid for the coupling of $\vpc$ to $\chi$ given by
expression (\ref{lphichi}). For other forms of couplings and for 
produced particles with non-zero spin 
the r.h.s. would have different forms (see below). The term describing the
back reaction of the produced particles is determined by the form of the 
interaction Lagrangian of the field $\vpc$ with the produced quantum fields,
and does not depend upon the potential $V(\vpc)$. This is in contrast to
the naive description of particle production  by the term $\Gamma \dot \vpc$ 
with a constant $\Gamma$, suggested in earlier literature. It
may approximately describe the realistic situation only for the harmonic
potential, $V=m^2 \vp^2 /2$. If one wants to mimic particle production
for other types of the potential by a similar local term, one would
immediately conclude that such a term must have different forms for 
different potentials.

The integral in the r.h.s. of eq. (\ref{scalres}) is logarithmically
divergent at $\tau \rightarrow 0$. This divergence is related to the mass
renormalization and can be regularized in the following way. Let us separate 
the integral in the r.h.s. into two parts, one from 0 to some $t_1$ and the
other from $t_1$ to $(t-t_{in})$. In the first integral let us subtract
and add $\vpc (t)$. We obtain:   
\be
&&\ddot{\vpc} + V'(\vpc) - 
 \frac{f^2}{4 \pi^2}\,\vpc \ln { t_1\over\epsilon } \nonumber \\
&&= \frac{f^2}{4 \pi^2} \int_0^{t_1} \frac{d \tau}{\tau} \, 
\left[ \vpc(t-\tau) - \vpc(t) \right]
+ \frac{f^2}{4 \pi^2} \int_{t_1}^{t-t_{in}} \frac{d \tau}{\tau} \, 
 \vpc(t-\tau)~. 
\label{renchi}
\ee
The integrals in the r.h.s. are now finite. 
The logarithmically infinite contribution in the l.h.s., related to the 
ultraviolet cut-off $\epsilon \rightarrow 0$, is taken out by the mass 
renormalization, so with a possible bare mass term in the potential,
$V_{m_0} = m_0^2 \vpc^2 /2$ (here $m_0$ is the bare mass of $\vpc$), we obtain:
\be
&&
\ddot{\vpc} + m^2 (t_1) \vpc + \left[ V'(\vpc) -V'_m (\vpc) \right] \nonumber \\
&&=
\frac{f^2}{4 \pi^2} \int_0^{t_1} \frac{d \tau}{\tau} \, 
\left[ \vpc(t-\tau) - \vpc(t) \right]
+ \frac{f^2}{4 \pi^2} \int_{t_1}^{t-t_{in}} \frac{d \tau}{\tau} \, 
 \vpc(t-\tau) ~,
\label{finchi}
\ee
where  $t_1$ is an arbitrary normalization point and the 
"running" mass is $m^2 (t_1) = m^2 (t_2) - (f^2/4\pi^2) \ln (t_1/t_2)$.
In eq.~(\ref{finchi}) we explicitly separated the massive part, $V_{m_0}$,
in the potential, so that the term in the square brackets vanishes for
the harmonic potential, $V(\vp) = m_0^2 \vp^2 /2$
  
In the limit of a small coupling, $f$, and for the harmonic potential,
equation (\ref{finchi}) can be solved analytically~\cite{df}. 
In this limit its solution coincides with that of the naive 
eq.~(\ref{eqgam}). We will look for the solution in the form:
\be
\vpc  = A(t) \cd \mbox{sin}(mt)~,
\label{phia}
\ee
where A is a slowly varying function of $t$. Substituting this into 
eq.~(\ref{finchi}) we obtain:
\be
-2 \dot{A} m \, \mbox{cos}(mt) = \frac{f^2}{4 \pi^2} \, \frac{\pi}{2} \,A \, 
 \mbox{cos}(mt)~,
\label{adot}
\ee
and correspondingly $A(t) \sim \exp (-\Gamma t/2)$ with $\Gamma = f^2/8\pi m$.
It is the correct decay width of $\vpc \rightarrow 2\chi$. Thus we found,
that in the limit of a weak coupling both equations (\ref{eqgam}) and 
(\ref{finchi}) have the same solution, though the equations themselves are
quite different. 

For the case of the harmonic potential one can reduce equation~(\ref{scalres})
to a more familiar form by making the Fourier transform:
\be
\vpc (t) = \int_{-\infty}^{\infty} {d\omega \over 2\pi} 
\exp (-i\omega t) \tilde \vpc (\omega)~,
\label{ftr}
\ee
where the spectral function $\tvpc (\omega)$ should be analytic in the
complex upper half-plane of $\omega$ to ensure vanishing of $\vpc$ for 
$t<t_{in}$ (below we take $t_{in}=0$); if $\vpc = {\rm const} \neq 0$
for $t<0$, this constant value can simply be subtracted from $\vpc$.
Assuming that $t_1$ is sufficiently small ($mt_1 \ll 1$), so that the 
integral from 0 to $t_1$ in the r.h.s. of eq. (\ref{renchi}) or (\ref{finchi}) 
can be neglected, we obtain:
\be
\tilde \vpc (\omega) \left( m_0^2 - 
{f^2 \over 4\pi^2} \ln {t_1 \over \epsilon} - \omega ^2 \right) = 
\frac{f^2}{4 \pi^2} \int^\infty_{-\infty} dt \,e^{i \omega t}\,
\int_{t_1}^{t} \frac{d \tau}{\tau} \, \vpc(t-\tau)~.
\label{fou}
\ee
For a large $t$ the integral over $d\tau$ depends weakly upon the upper limit,
so that it can be pushed to infinity and the integration over $t$ can be
done explicitly:
\be
\tilde \vpc (\omega) \left( m_0^2 - 
{f^2 \over 4\pi^2} \ln {t_1 \over \epsilon} - \omega ^2 \right) = 
\frac{f^2}{4 \pi^2} \tvpc \int_{t_1}^{\infty} \frac{d \tau}{\tau} 
\left( \cos \omega \tau +i\sin  \omega \tau \right)~.
\label{fou1}
\ee
The second integral in the r.h.s. is equal to 
$i\,m\Gamma \,{\rm sign} (\omega) \tvpc (\omega)$, while the first one,
logarithmically divergent in the lower limit, gives mass renormalization in the
momentum space. With the account of this term the renormalized mass can be 
written as:
\be
m_0^2 -{f^2 \over 4\pi^2} \ln {t_1 \over \epsilon} -
\frac{f^2}{4 \pi^2} \int_{\omega t_1}^{\infty} \frac{d y}{y} \cos y =
m_0^2-{f^2 \over 4\pi^2} \ln {\Lambda \over \omega } + ({\rm finite\,\,\,
terms}) \equiv m^2_{ren}~,
\label{renmass}
\ee
where $\Lambda = 1/\epsilon$ is the ultraviolet cut-off. 
Finally we obtain:
\be
\tilde \vpc (\omega) \left[ m_{ren}^2  
 - \omega ^2  -i\, m_{ren} \Gamma\, {\rm sign}(\omega) \right] = 0~.
\label{foufin}
\ee
This is a simple linear equation in momentum space. A similar equation
for an unstable particle moving in an external potential (gravitational
field) has been derived in ref. \cite{ad1}. Eq. (\ref{foufin}) is 
different from eq. (\ref{eqgam}) with the $\Gamma \dot \vpc$-term, but
in the limit of small $\Gamma /m$ it has the same solution (\ref{solgam}). 
With an increasing
$\Gamma /m$ the solution of eq.~(\ref{eqgam}) ceases to oscillate and turns
into an exponentially decaying one. For an even larger ratio $\Gamma /m$,
especially in the unphysical limit $\Gamma \gg m$, the solution of 
eq. (\ref{eqgam}) tends to a constant. The
solutions of eq. (\ref{foufin}) behaves differently. 
For any value of the ratio $\Gamma /m$ the
solution of this equation has both oscillating and exponentially decreasing
factors. This behavior is demonstrated by the numerical solution
of eq. (\ref{finchi}), see figs.~1.
Recall that eq. (\ref{finchi}) is equivalent to the algebraic one 
(\ref{foufin}) for the Fourier transformed field amplitude, only in the limit 
of a large $mt$. In the general case eq. (\ref{renchi}) or 
(\ref{finchi}) should be solved numerically. 
 
There is a subtle point related to the definition of the value of the
renormalized mass. It is clear that $m_{ren}$ in eq.~(\ref{foufin})
and $m$ in eq.~(\ref{solgam}) (correspondingly $m$ in $V(\vpc)$ in 
eq.~(\ref{eqgam})) are normalized at the same point and should be taken equal
when one compares the solution of these equations. On the other hand, the
value $m^2(t_1)$ in eq.~(\ref{finchi}) is different from $m^2_{ren}$,
because of a finite contribution to $m^2$ from the r.h.s. of this equation,
and depends therefore on $f$. This explains the presence of the coefficient
$\alpha \neq 1$ in the envelope exp$(-\alpha \Gamma t /2 )$ to
the numerical solution of eq.~(\ref{finchi}), presented in figs.~1.

There is no simple way to find an analytic solution for more complicated 
potentials, when the non-perturbed equation for $\vpc (t)$ is non-linear. 
For the case when the potential is dominated by the quartic term, 
$V(\vp) =\lambda_4 \vp^4 /4$, the equation  can be written as:
\be
z'' + z^3 = h^2 \int _{x_1}^{x} {dy\over y} z(x-y)~.
\label{zx}
\ee
Here the effective coupling constant is
$h^2 = f^2 /\left( 4\pi^2 \lambda_4 \vp_0^2 \right)$.
We introduced the dimensionless quantities $x=\sqrt \lambda_4 \vp_0 t$
and $z(x) = \vp /\vp_0$, where $\vp_0$ is the characteristic magnitude of 
the field $\vp$. Prime means differentiation with respect to $x$.
The contribution to the integral from 0 to $x_1$ is
neglected as in the previous case (but it can easily be taken into account).
We will solve this equation for $\vp (t=0) = \vp_0$ and 
$\dot \vp (t=0) =0$.  

In the limit of $h=0$ this equation is solved in terms of Jacobi elliptic
functions (see e.g. \cite{gr}):
\be
z_0 (x) = {\rm cn} (x,1/\sqrt 2)~.
\label{z0}
\ee
In what follows we will use the notation 
${\rm cn} (x, 1/\sqrt 2) \equiv {\rm cn} (x)$.
It is easy to find the first order corrections in $h^2$ in the limit of 
relatively small time, $h^2 x \ll 1$. We will look for the solution in the 
form:
\be
z(x) = A(x) \, {\rm cn } \left[\beta(x) \right]~,
\label{zx1}
\ee
where $A(x)$ is a slowly varying function of $x$ and 
$\beta'(x) = A(x)$, which ensures vanishing of the lowest order
terms in the l.h.s. of eq. (\ref{zx}). The function $A(x)$  differs 
from the one used above by the constant factor $\vp_0$. Substituting 
expression (\ref{zx1}) into eq. (\ref{zx}) and neglecting the term 
proportional to $A''$ we obtain:
\be
3A'A {d \over d\beta}\left[{\rm cn}\, \beta(x)\right]
= h^2 \int^x {dy\over y} A(x-y) {\rm cn}\left[ \beta (x-y) \right]~.
\label{a'a2}
\ee
In the first order in $h^2 x$ the solution is straightforward: 
\be
A(x) \approx 1 - 0.62 h^2 x 
\approx 1-0.016 f^2\left(\sqrt{\lambda_4} \vp_0\right)^{-1} t~.
\label{ax}
\ee
Unfortunately we failed to find an analytic 
approximation to the asymptotic decrease of the amplitude of the oscillating 
field $\vp$ at very large $x$,
which characterizes the particle production rate at a large time. The problem 
is that $A(x)\rar 0$ for $x\rar \infty$ and $\beta (x) \rar {\rm const}$.
Correspondingly the decomposition of $z$ as a product of quickly and slowly
varying factors becomes invalid.
However, the numerical treatment of the problem is quite simple and 
straightforward. 
The r.h.s. of equation (\ref{zx}) gives a nonzero contribution into
the mass of the field $\vp$, as is easy to see a negative one, 
$\delta m^2 < 0$. Correspondingly the solution, found numerically, 
asymptotically tends to the nonzero value, $\vp^2_{fin} =  
-\delta m^2 /\lambda$. A counter-term $\delta m^2 \vp^2/2$ can be added to
the potential, $V(\vp)$, to ensure vanishing of the renormalized mass.
The equation is solved in this case too and, as expected, 
$\vp \rar 0$. The decrease of its amplitude at small $h^2 x$ agrees quite 
well with the approximate result (\ref{ax}) as is seen 
in figs.~2. 

In connection with the derivation of the equation of motion for the
primary field $\vp$, given above and in the following sections,
a very important question may arise.
We derived equation (\ref{finchi}) perturbatively including only one-loop
contribution into $\langle \chi \chi \rangle$. 
Could one go beyond perturbation 
theory using this equation? We do not have a rigorous answer to this question
(see also the discussion in the last section).
However, there are quite many examples,  that an equation with a perturbative
potential (for example the Schroedinger equation with Coulomb potential)
permits to go beyond perturbation theory. Moreover,  the fact that equations 
(\ref{eqgam}) and (\ref{finchi}) are different even in perturbation 
theory have important implications~\cite{df,dfrs} e.g. for spontaneous 
baryogenesis \cite{ck,ad}.

\section{Quartic scalar coupling}

The calculations are essentially the same for other types of coupling of 
$\vpc$ to quantum fields. For the interaction of the form
${\cal L}_{int} = \lambda_2 \vpc^2 \chi^2$ the equation of motion for $\vpc$
takes the form:
\be
 \ddot{\vpc} + V'(\vpc) = \frac{\lambda^2_2}{2 \pi^2} \, \vpc(t) \,
\int_0^{t-t_{in}} \frac{d \tau}{\tau} \,\vpc ^2 (t - \tau)~.
\label{vp2chi2}
\ee
The logarithmic divergence at $\tau \rightarrow 0$ can be removed by
renormalization of the coupling constant, $\lambda_4$, in the self-interaction
potential, $V_\lambda(\vpc)=\lambda_4 \vpc^4/4$. It can be done exactly 
in the same way as it 
was done above for mass renormalization. We separate the integral in the
r.h.s. of eq. (\ref{vp2chi2}) into two parts: from 0 to $t_1$ and from $t_1$
to $(t-t_{in})$. In the first integral we subtract and add $\vpc^2 (t)$ and
obtain:
\be
 \ddot{\vpc} + \lambda_4 (t_1) \vpc^3 + \left[ V'(\vpc)- 
V'_\lambda(\vpc) \right] = 
\frac{\lambda^2_2}{2 \pi^2} \, \vpc(t) \,
\int_0^{t_1} \frac{d \tau}{\tau} \left[ \vpc ^2 (t - \tau)-\vpc^2 (t)
\right] \nonumber \\
+\frac{\lambda^2_2}{2 \pi^2} \, \vpc(t) \,
\int_{t_1}^{t-t_{in}} \frac{d \tau}{\tau} \vpc ^2 (t - \tau)~,
\label{lam}
\ee
where $\lambda_4 (t_1)$ is the coupling constant of the 
$\vp^4$ self-interaction,
renormalized  at the normalization point $t_1$,
by the loop with the field $\chi$,  $\lambda_4 (t_1)=
\lambda_4 (t_2) - (\lambda_2^2 /2\pi^2) \ln (t_1/t_2)$. The integrals in the
r.h.s. are  now finite. We assume that $t_1$ 
is small, $mt_1 < 1$, while the solutions will be taken in the limit
$mt>1$.  Here we neglect the loop with the field $\vp$ itself or, in
other words, neglect the self-production of $\vp$-quanta by the classical
field $\vp_c$. This is physically sensible if $\lambda_4 < \lambda_2$. 

A perturbative solution in this case is slightly more complicated than that  
given in the previous section.
We assume that the potential is dominated by the harmonic part, 
$V(\vpc) \approx m^2\vpc^2 /2$, and that the coupling constants are small,
$\lambda_4 < \lambda_2 < 1$. To the lowest order in the coupling constants
the solution has the form:
\be
\vpc (t) = A(t) \cos (mt+\Phi) +  A^3 (t) \left[
 c_1 t \sin (mt+\Phi) +c_2 \sin (3mt+3\Phi) + 
c_3 \cos (3mt+3\Phi) \right]~,
\label{philam}
\ee
where $\Phi$ is a constant phase determined by initial conditions.
The slowly varying amplitude $A(t)$ and the coefficients $c_j$ are 
given by:
\be
A(t) = A_{in} \left[ 
1+ {\lambda_2^2 \over 16\pi}{ A_{in}^2\over m} (t-t_{in}) \right]^{-1/2},
\label{alam}
\ee
\be
c_1 = - {3\bar \lambda \over 8m} + 
{\lambda^2_2\over 4\pi^2 m} \int_0^{mt} {d\eta\over \eta} \sin^2 \eta~,
\label{c1}
\ee
\be
c_2 = -\lambda_2^2/(128\pi m^2)~,
\label{c2}
\ee
\be
c_3 = \bar \lambda /(32 m^2)~,
\label{c3}
\ee
where:
$$
\bar \lambda = \lambda_4 - 
{\lambda^2_2\over 2\pi^2} \int_{mt_1}^\infty {d\eta \over \eta} \cos 2\eta~.
$$
The term proportional to $c_1$ in equation (\ref{philam}) rises linearly with
time (in addition $c_1$ itself rises as $\ln mt$). This linear rise is the
well known resonance behavior associated with the  
equality of the eigen-frequency of the
non-perturbed equation and the perturbative force, coming both from the term
proportional to $\lambda_4 \vpc^3$ and from the terms in the r.h.s.
of eq.~(\ref{lam}). Moreover, if $\lambda_4$ is non-negligible (to be more
precise, the combination $\lambda_4 A^2_{in} /m^2$), then the term 
$\lambda_4 \vp_0^2 \vp_1 $ (where $\vp_0$ is the zero-order approximation
and $\vp_1$ is the first-order correction) would induce a parametric
resonance and an exponential rise of $\vp$. Of course these resonances
do not have physical sense, because the original equation does
not possess any instability. The "resonances" (both the usual and the 
parametric ones) are quickly turned off with rising time and the large 
time behavior remains non-resonant, as is confirmed by the numerical 
calculations. A resonance amplification of the boson production rate would
take place if the accumulation of the produced bosons is taken into account. 
To do that one has to make quantum averaging of the equations of motion,  
not over the vacuum state, as is done here, but over the 
states with non-zero boson occupation number. It will be done in
a subsequent work~\cite{dh}.

The amplitude of the field, $\vpc (t)$, decreases in accordance with 
expression (\ref{alam}). This decrease is much slower than the exponential
decay of $\vpc (t)$ considered in the previous section. Correspondingly
the efficiency of particle production in the case of the quartic coupling 
is much weaker than that induced by the trilinear coupling in the
previous section. The usual description of the back reaction of
particle production by the $\Gamma  \dot \vpc$ term in this case is very 
far from reality. Correspondingly the approach of the earlier 
papers~\cite{mst,st} (see also the recent review~\cite{laz}) should be
modified.
The results of the 
numerical solution of eq. (\ref{lam}) are presented in figs.~3.
For a weak coupling, i.e. for $\lambda^2_2 m^2 /2\pi^2 A_{in}^2 \ll 1$, 
they agree quite  well with the perturbative result~(\ref{philam}). The 
fitting coefficient $\alpha \neq 1$ appears because of renormalization
of the parameters entering the l.h.s. of eq. (\ref{lam}) due to the 
interaction term in the r.h.s., while the dependence on time remains
the same as that given by eq. (\ref{alam}).

The case of the quartic potential, $V(\vp) = \lambda_4 \vp^4 /4$, can be
treated similarly to that in the previous section. The solution is searched 
for in the form (\ref{zx1}), so that the l.h.s. of eq.(\ref{a'a2}) remains the 
same, but the r.h.s. becomes different:
\be
3A'A {d\over d\beta}{\rm cn}\left[ \beta(x)\right] 
={\lambda_2^2 \over 2\pi^2 \lambda_4} {\rm cn}\left[ \beta (x) \right]
\int^x {dy\over y} A^2(x-y) {\rm cn^2}\left( \beta (x-y) \right)~,
\label{a'a2lam2}
\ee
where $x= \sqrt{\lambda_4} \vp_0 t$.

The decrease of the 
amplitude of the oscillations is described by the expression:
\be
A(t) \approx {1\over 1+ 0.0085 (\lambda^2_2 /\lambda_4) \vp_0 t }~.
\label{afin}
\ee
To obtain this result, only the terms $\cos (px)$ and $\cos (3px)$ (with 
$ p \approx 0.85$) in the Fourier decomposition of 
cn$\left[ \beta (x) \right]$ (see e.g. \cite{gr}) were taken into 
account. Again this behavior agrees quite well with the numerical 
solution for small couplings, see figs.~4. As above the fitting parameter
$\alpha \neq 1$ is induced by renormalization. 

It is interesting to note, that  for the quartic coupling considered 
in this section the oscillations asymptotically decrease faster in the 
case of the $\vp^4$-potential, eq. (\ref{afin}), than in the case 
of the $\vp^2$-potential, eq. (\ref{alam}). Further, comparing figs.~1 and 2 
with figs.~3 and 4 (and the approximate behaviour of the envelopes) 
one realizes,
that for the trilinear coupling, $f \vp \chi^2$, the scalar field, $\vp$,
disappears much faster than for the quartic coupling. This is what one
would expect, since the probability of the reaction induced by the
coupling $\lambda_2 \vp^2 \chi^2$ is quadratic
in the $\vp$-number density, whereas for the trilinear coupling the 
probability is linear in the $\vp$-number density.
 
\section{Coupling to fermions }

When the classical scalar field couples to quantum fermions the
calculations are slightly more complicated, but still straightforward.
We will consider the Lagrangian density where the classical scalar field,
$\vpc$, couples to the
spin $1/2$ field $\psi$ through the Yukawa coupling:
\[
{\cal L}(\vp , \psi, \overline \psi) =  
\frac{1}{2} (\partial \vp)^2 - V(\vp) + i \overline{\psi}
\gamma_\mu\partial^\mu \psi + {g} \overline{\psi} \psi \vp~,
\]
where $\psi$ is quantized according to:
\beq
\psi (x) = \int \frac{d^3p}{(2 \pi )^3 \sqrt{2 E}} \, \sum_s
\left( a^s_p u^s(p) e^{-ipx}  + b^{s \dag}_p v^s(p) e^{ipx}\right)~,
\label{qmferm}
\eeq
where $a$ and $b^\dag$ are annihilation and creation operators 
at momentum $p$ and spin $s$, obeying the anti-commutation relations:
\[
\{ a^r_p , a^{s \dag}_q \} = \{  b^r_p , b^{s \dag}_q \} = 
(2 \pi)^3 \delta^3(\vec{p}-\vec{q}) \delta^{r s}~.
\]
Summation over spins, which we need in what follows, is achieved 
with the usual relations:
\[ 
\sum u^s(p) \overline{u}^s(p) = \slash \! \! \!p  + m \,\,\,\,\,   \mbox{and}
\,\,\,\,\,  
\sum v^s(p) \overline{v}^s(p) = \slash \! \! \!p  - m.
\]
The equations of motion are:
\be
\ddot{\vp} -\Delta\, \vp + V'(\vp) = -{g} \overline{\psi} \psi~,
\label{fermvp} \\
i \slash \! \! \! \partial \psi + {g} \psi \vp =0~,
\label{fermpsi} \\
-i \partial_\mu (\overline{\psi} \gamma^\mu) +  {g} 
\overline{\psi} \vp =0 ~.
\label{fermpsibar}
\ee
In the background of the classical field $\vpc(t)$ 
eq.~(\ref{fermpsi}) is formally solved by:
\beq
\psi(x) = \psi_o(x) + i {g} \int d^4 y \, G^R_f(x,y) \, \psi(y) \vpc (y)~,
\label{solferm}
\eeq
where $\psi_0$ is an initially free fermion field and 
the retarded Green's function for fermions is given by:
\[
G^R_f(x,y) = i\int \frac{d^4p}{(2 \pi)^4} \, \frac{\slash \! \! \!p 
+ m}{p^2 - m^2} 
\, e^{-ip\cd(x-y)}~.
\]

Here we will consider the case of massless fermions, which is 
technically simpler, and take fermionic masses into account in a
subsequent paper~\cite{dh}. There are some subtleties in the calculations
in comparison with the case considered above of production of massless scalars,
especially regarding renormalization and regularization. However, 
the calculations can essentially be reduced to the scalar case.
The fermionic Green's function is expressed through the scalar one as:
\be
G^R_f(x,y) =  i \slash \! \! \! \partial_x G^R_s(x-y)~.
\label{grf}
\ee
The vacuum expectation value of the field operators in the r.h.s. of 
eq.~(\ref{fermvp}) is equal to:
\be
g\langle \bar \psi (x) \psi (x) \rangle =
g^2 \int d^4 y \vp_c (y)  {\partial G^R_s (x-y) \over \partial x^\mu }
{\rm Tr} \langle  \bar \psi_0 (x) \gamma_\mu \psi_0 (x) \rangle~.
\label{rhsf}
\ee
Using the representation (\ref{qmferm}) and taking vacuum expectation value 
and spinor trace we obtain:
\be
 g\langle \bar \psi (x) \psi (x) \rangle  =
4g^2 \int d^4 y \,\vp_c (y) \, {\partial G^R_s (x-y) \over \partial x^\mu }
\,{\partial S(x-y) \over \partial x^\mu }~,
\label{rhsf1}
\ee
where:
\be
S(x-y) = \int {d^3 k \over (2\pi)^3 2E } e^{ik(x-y)}~,
\label{s}
\ee
is the same function (coming from vacuum averaging of free field operators)
which enters the r.h.s. in the scalar case. It satisfies the condition:
\be 
\partial^2 S = 0~,
\label{d2s}
\ee
while the Green's function satisfies:
\be 
\partial^2 G(x-y) = \delta^4 (x-y)~.
\label{d2del}
\ee
Using these two equations and the identity:
\be
2 \partial_\mu S \partial_\mu G = \partial^2 (SG) - G\partial^2 S
-S\partial^2 G ~,
\label{d2sg}
\ee
we obtain:
\begin{eqnarray}
 \ddot{\vpc} + V'(\vpc)  = 
-{g^2 \over 4\pi^2} {d^2\over dt^2}  \int_0^{t-t_{in}} 
\frac{d \tau}{\tau} \, \vpc(t-\tau) - g^2 \Lambda ^2 
\vpc~.
\label{eqnferm}
\end{eqnarray}
The last term, proportional to the ultraviolet cut-off parameter, $\Lambda$,
renormalizes the mass of $\vpc$. The term under the integral sign is the
same as the one in the scalar case eq. (\ref{scalres}) and its logarithmic
divergence at $\tau =0$ can be treated in a similar way. However, instead of
the term $\vp_c \ln (t_1/\epsilon)$ we obtain 
$\ddot \vp_c \ln (t_1/\epsilon)$. This corresponds to wave function
renormalization (renormalization of the 
kinetic term) at the time moment $t_1$.
Finally we obtain:
\be
&& \left[1+ {g^2\over 4\pi^2}
\ln \left({t_1\over\epsilon} \right)\right]\ddot{\vpc} + V'_{ren}(\vpc) 
\nonumber \\
&& =
-{g^2\over 4\pi^2} 
{d^2\over dt^2} \left\{ \int_{0}^{t_{1}} \frac{d \tau}{\tau} \, 
\left[ \vpc(t-\tau) - \vpc(t) \right]
+  \int_{t_1}^{t-t_{in}} \frac{d \tau}{\tau} \, 
 \vpc(t-\tau) \right\}~,
\label{ferfin}
\ee 
where $V_{ren}(\vpc) $ is the potential of $\vpc$ with renormalized mass.

The perturbative solution in the case of a harmonic potential, 
$V(\vpc ) = m^2\vpc^2/2$, is essentially the same as in section~\ref{tril}.
If we assume $\vpc = A(t) \, \mbox{sin}(m t)$,  where $A(t)$ is a slowly
varying function of $t$, we find that 
$\dot{A}/A = g^2 \, m/ (16 \pi)$. For a small $\Gamma$ it is essentially 
the same result as 
can be found from the naive ansatz with the $\Gamma \dot{\vpc}$ term 
in the r.h.s.
of eq.~(\ref{eqgam}) using
$\Gamma_{\vpc \overline{\psi} \psi} = g^2 m_{\vpc}/(8 \pi)$~\cite{df}.
For the harmonic potential, $V(\vp) = m^2\vp^2 /2$, and the Yukakwa coupling
to fermions, $g\vp \bar \psi \psi$, the particle production is quite
similar to that considered in the first part of section 2. The results of
the numerical calculations for this case are presented in figs.~5.

For the quartic potential, $V(\vp) = \lambda_4 \vp^4 /4$, we will look for
the perturbative solution in the form (\ref{zx1}). In direct analogy with the
previous sections it is easy to find, that the amplitude $A(x)$ decays as:
\be
A(t) \approx {\vp_0  \over 1 + 0.011 g^2 \sqrt{\lambda_4} \vp_0 t }~.
\label{at2}
\ee
The results of the numerical calculations are presented in figs.~6.

Now, one could also consider non-renormalizable couplings, such as
${\cal L}_{int} = \kappa \vp^2 \bar{\psi} \psi$, where $\kappa$
has dimension (mass)$^{-1}$. The corresponding equation of motion has the 
form:
\be
\ddot{\vpc} + V'(\vpc)  = 
-{\kappa^2 \over 2 \pi^2} \, \vp (t) \, {d^2\over dt^2}  \int_0^{t-t_{in}} 
\frac{d \tau}{\tau} \, \vpc(t-\tau)^2~.
\label{nonren}
\ee
An approximate analytic solution for the potential 
$V(\vp) = m^2\vp^2 /2$ can be found
along the same lines as  presented in the beginning of section 3.
The essential part of the solution behaves as:
\be
\vp (t) = \vp_0 \left( 1 + { \vp_0^2 \kappa^2 \over 4\pi} mt \right)^{-1/2}
\cos mt~.
\label{phikap}
\ee
This result describes the numerical solution quite well (see figs.~7). 

If the potential is dominated by the $\lambda_4 \vp^4 $-term, 
eq.~(\ref{nonren}) 
can be solved with the ansatz (\ref{zx1}). The amplitude decays as:
\be
A(t) = \left( 1 + 0.075 \sqrt{\lambda_4} \kappa^2 \vp_0^3 t 
\right)^{-1/3}
\label{kaplam}
\ee
The decay is astonishingly slow. It can be compared with the numerical
solutions presented in figs.~8. The agreement is quite good if the above
mentioned effect of renormalization is taken into account.

The numerical solution of eq.~(\ref{nonren}) can be compared to the
solution of eq.~(\ref{ferfin}) and
one sees  that, as in the bosonic case, the scalar field, $\vp (t)$, 
falls off much
faster for the coupling $g \vp \bar{\psi} \psi$ than for
$\kappa \vp^2 \bar{\psi} \psi$. The reason being, that the latter 
goes quadratically with the number density.

\section{Discussion and conclusion}

We have seen that with a weakly coupled  external field, $\vpc (t)$, the
effect of particle production on its evolution can be described by
the addition of an extra term into the usual Klein-Gordon equation. This
extra term is given by an integral over all previous history of the 
field, $\vpc (t)$, and the concrete form of the integrand is determined by
the form of the interaction of $\vpc$ with the produced particles (see examples
considered above). In all the cases it has been assumed that the interaction 
is bilinear in terms of the produced quantum fields:
\be
{\cal L}_{int} = F_b (\vp) \chi^* \chi + F_f (\vp) \bar \psi \psi~,
\label{bf}
\ee
where $\chi$ and $\psi$ are bosonic and fermionic quantum fields respectively.
The calculations have been done in the limit of a negligible mass of the 
produced particles and in flat space-time. Lifting both restrictions
is straightforward but tedious and will be considered elsewhere \cite{dh}.

The most essential restriction is the weakness 
of the field, $\vpc (t)$, such that 
the Green's functions of the produced particles can be 
taken as the free field ones.
This assumption is justified if the effective mass of the produced quanta,
generated by the interaction (\ref{bf}), is small in comparison with their
characteristic energy. The latter is essentially the frequency of the 
oscillations of $\vpc (t)$. In this approximation our equation
adequately describes the decrease of the amplitude of $\vpc (t)$ induced by
the particle production.

Another simplifying assumption, which was made above, 
is that the number density of the 
produced particles is small, so that neither Fermi-Dirac suppression nor
Bose-Einstein enhancement is essential. Correspondingly the production of 
fermions would be further suppressed in comparison with our results, while 
the production of bosons could be quite significantly enhanced. The effect of 
quantum statistics is expected to give rise to a resonance 
amplification of the boson production and will be considered elsewhere.
Finally we assume, that $\lambda_4$ in the potential is so small 
that we can ignore self-production of $\phi$-quanta by the field $\vpc$.

The case of a strong external field is considerably more complicated. There
is no closed expression for the Green's function of the produced particles
in a strong external field and no closed equation for $\vpc (t)$ can be
written down. To overcome this problem one may use an adiabatic approach.
In the case of the bilinear coupling to the produced particles (\ref{bf}) 
the calculation of the particle production rate can be reduced to the
solution of an ordinary differential equation with time-dependent
frequency (see e.g. \cite{gmm}). It is all done in the "fixed" field 
approximation i.e. without field decay due to particle production. The 
corresponding decrease of the field amplitude can be taken into account
adiabatically by imposing conservation of the total energy.
In the historically first papers \cite{dl,afw,astw} on particle production 
by the inflaton field, the process was considered perturbatively in a spirit 
close to the present work, but without inclusion of 
the effect of back-reaction of the 
produced particles on the evolution of the inflaton field.

The non-perturbative calculations, valid for an arbitrary strongly
coupled field, $\vpc (t)$, were first undertaken in refs.~\cite{dk,tb}. 
In both papers the possibility of an
enhanced production of bosons due to parametric resonance excitation was
noticed, but it was argued~\cite{dk} that parametric resonance is
not operative, in the framework of the model considered there, because of the 
cosmological expansion and rescattering of the produced particles. 
It has later been found that the resonance may be wide and the effects
of particles leaving the resonance mode is not so important,
permitting parametric enhancement of 
boson production by the inflaton decay (see ref.~\cite{kls}). 

Non-perturbative production of fermions was considered in ref.~\cite{dk} for
the usual Yukawa coupling to fermions, $g \vp \bar \psi \psi$, and for
an arbitrary time dependence of the scalar field, $\vpc (t)$. Concrete
examples were given for the case of harmonic oscillations of $\vp$, 
which take place if $\vp$ "live" in the potential, $V(\vp)=m^2\vp^2 /2$,
and the back reaction of particle production on $\vp$-evolution is neglected.
It was shown there that in the weak field limit the rate of particle
production is described by a constant decay width, $\Gamma$, in accordance
with perturbation theory. In the limit of a strong field the production 
probability goes down as $1/\vp_0^{1/2}$, where $\vp_0$ is the amplitude of
the oscillations of the field $\vp (t)$. This suppression is related to an
increase of the effective mass of the produced fermions, $m_{eff}\sim g\vpc$,
while the frequency of the oscillations of $\vp$ remains constant.
This effect, of course, cannot be traced in the present work, where the
weak field approximation is used.
 
In a recent paper \cite{bhp} a non-peturbative approach has been applied to 
fermion production by the field $\vpc (t)$ with the potential 
$m^2\vp^2/2+\lambda \vp^4/4!$. A system of coupled equations for the
evolution of the field, $\vpc$, and the mode functions of the produced 
fermions was solved numerically in one loop approximation. This permitted to
take the back reaction of particle production on
$\vpc$-evolution into account. 
However, our equation differs from that derived in 
ref.~\cite{bhp} in the weak field limit (linear in the field $\vpc$).
This difference can possibly be ascribed to a difference in our approaches
to renormalization and the initial time singularity. As is seen, from the
arguments presented here, we treat the initial time singularity as a usual
ultraviolet singularity which can be regularized by the standard
renormalization procedure. In ref.~\cite{bhp} this singularity was avoided 
with the help of a Bogoliubov transformation. In the limit of a large
field strength, such that the effective mass of fermions, $g\vpc$, is larger
than the mass of the field $\vpc$, the calculations of ref.~\cite{bhp}
showed a strong suppression of particle production in agreement with 
ref.~\cite{dk}. However, it is found there that the suppression  persists
even in the case of $g\vpc \ll m_{\vp}$, though not for extremely small 
$g\vpc$. This result is in contradiction with the 
calculations of ref.~\cite{dk},
according to which the production rate in this region of parameters is
well described by perturbation theory. 

Recently the approach of the paper~\cite{dk} was repeated in ref.~\cite{gk} 
for the particular case of particle production by the scalar field in the 
potential $\lambda \phi^4$. It was found that the suppression
of the particle 
production at large values of the field amplitude for the case of
harmonic potential, as observed in~\cite{dk}, does not take place for 
$V=\lambda \phi^4 /4$, and the 
production of fermions may be quite efficient. This conclusion was also made
(a little earlier) in ref.~\cite{bhp}.    
Possible explanations of the
deviation of these results, from the results of ref.~\cite{dk}, is an 
influence of the $\lambda \vp^4$-term in the potential and corresponding
excitation of higher modes, as well as an increase of 
the oscillation frequency of $\vpc$ due to a bigger amplitude of $\vp$.

Returning to the results of the present paper, we repeat that they are valid
in the weak field approximation and, as such, can be applied to the
universe heating by the inflaton decay at the final stage of the inflationary
process, when the amplitude of the inflaton field is sufficiently small. 
Even though the inflaton energy during
this period only contributes a relatively small fraction of the
initial inflaton energy, its role in the creation of the matter in the
universe could be much more significant. The reason for that is that at 
the final stage of the decay, when the potential is approximately 
$m^2\vpc^2/2$, the inflaton behaves as non-relativistic matter. Therefore 
its energy density is amplified, with respect to the energy of the 
relativistic particles produced earlier, by the red-shift factor 
$a(t)/a_{in} (t)$. This makes the role of the final part of the inflaton decay 
correspondingly enhanced and the temperature at the end of the
decay becomes close to the lowest order perturbative
result~\cite{fkyy}. Another mechanism of suppression of particle production
was discussed in ref.~\cite{ac}, where it was shown that parametric
amplification might be effectively suppressed by the final state
interaction of the produced particles.

Another possible application of our results is spontaneous baryogenesis.
The evolution of the spatially homogeneous pseudo-goldstone field, $\theta$,
which is related to spontaneous
breaking of baryonic charge conservation, is described by the equation:
\be
f^2 \ddot \theta + U'(\theta) = {d\over dt} J_B^0 ~,
\label{theta}
\ee
where $J_B^0$ is the baryonic charge density and $f$ is the scale of 
spontaneous symmetry breaking.
Since, as we have shown, the impact of particle production on the evolution 
of $\vpc$ is not given by the term $\Gamma \dot \vp$, it means that the
naive identification $f^2\Gamma \dot\theta = d J_B^0 /dt$ is not correct.
Correspondingly the
baryon asymmetry is not given by $\delta J_B^0  \sim \Gamma \delta \theta$
(for a more detailed discussion see refs.~\cite{df,dfrs}).

{\bf Acknowledgment.}
The work of AD and SH was supported in part by the Danish National Science 
Research Council through grant 11-9640-1 and in part by Danmarks 
Grundforskningsfond through its support of the Theoretical Astrophysical 
Center. AD is greatful to the Scuola Normale Superiore (Pisa) where a part
of this work has been done.

\newpage

{\large \bf Figure Captions:}
\vskip0cm

\noindent
{\bf Fig. 1} $~~~$ 
Trilinear scalar coupling, $f \vp \chi^2$, and harmonic potential, 
$V=m^2 \phi^2 /2$. The decrease of the amplitude is given by the envelope
function, exp$(-\alpha \Gamma t/2)$, $\Gamma = f^2/8 \pi m$.
Fig.~1a: $m =1, f=0.7, \lambda = 0, \alpha = 1.03$.
Fig.~1b: $m =1, f=1.4, \lambda = 0, \alpha = 1.25$.

\vskip0cm
\noindent
{\bf Fig. 2} $~~~$ 
Trilinear scalar coupling, $f \vp \chi^2$, and potential, 
$V=\lambda \phi^4 /4$. 
The decrease of the amplitude is given by the envelope
function, $1-0.016 \alpha f^2 (t - t_{in})$.
Fig.~2a: $m_r =0, f=0.5, \lambda =1, \alpha = 0.94$.
Fig.~2b: $m_r =0, f=1.0, \lambda =1, \alpha = 0.92$.

\vskip0cm
\noindent
{\bf Fig. 3} $~~~$ 
Quartic scalar coupling, $\lambda_2 \vp^2 \chi^2$, and harmonic potential, 
$V=m^2 \phi^2 /2$. The decrease of the amplitude is given by the envelope
function, 
$\vp _0 \, \left( 1+\alpha \lambda_2^2/(16 \pi) 
\vp_0^2/m (t-t_{in}) \right) ^{-1/2}$.
Fig.~3a: $m = 1, \lambda_2 =0.5, \lambda_4 =0, \alpha = 1$.
Fig.~3b: $m = 1, \lambda_2 =1.5, \lambda_4 =0, \alpha = 1.5$.

\vskip0cm
\noindent
{\bf Fig. 4} $~~~$ 
Quartic scalar coupling, $\lambda_2 \vp^2 \chi^2$, and potential, 
$V=\lambda_4 \phi^4 /4$. The decrease of the amplitude is given by 
the envelope
function, 
$\vp _0 /(1+ 0.085 \alpha \lambda_2^2/\lambda_4 \vp_0 (t-t_{in}))$.
Fig.~4a: $m_r =0, \lambda_2 =0.5, \lambda_4 =1, \alpha= 1.05$.
Fig.~4b: $m_r =0, \lambda_2 =1.0, \lambda_4 =1, \alpha = 1.3$.

\vskip0cm
\noindent
{\bf Fig. 5} $~~~$ 
Trilinear fermion coupling, $g \bar \psi \psi \vp$, and harmonic potential, 
$V=m^2 \phi^2 /2$. The decrease of the amplitude is given by the envelope
function, exp$(-\alpha \Gamma t/2 )$, $\Gamma = g^2 m/(8 \pi)$.
Fig.~5a: $m=1, g=0.5, \lambda =0, \alpha = 1.0$.
Fig.~5b: $m=1, g=1.0, \lambda =0, \alpha = 0.8$.

\vskip0cm
\noindent
{\bf Fig. 6} $~~~$ 
Trilinear fermion coupling, $g \bar \psi \psi \vp$, and potential, 
$V=\lambda_4 \phi^4 /4$. The decrease of the amplitude is given by the envelope
function, $\vp_0 /(1+ 0.011 \alpha g^2 \sqrt{\lambda_4} \vp_0 (t-t_{in}))$.
Fig.~6a: $m=0, g=0.5, \lambda =1, \alpha =1.0$.
Fig.~6b: $m=0, g=1.0, \lambda =1, \alpha =0.82$.

\vskip0cm 
\noindent
{\bf Fig. 7}
Quartic fermion coupling,  $\kappa \vp^2 \bar{\psi} \psi$, 
and harmonic potential, 
$V=m^2 \phi^2 /2$. The decrease of the amplitude is given by the envelope
function, $ \left[ 1 + \alpha \vp_0^2  \kappa^2 mt /(4\pi)\right]^{-1/2}$.
Fig.~7a: $m=1, \kappa =0.5, \lambda=0, \alpha = 1.0$.
Fig.~7b: $m=1, \kappa =1.0, \lambda=0, \alpha = 1.0$.

\vskip0cm        
\noindent
{\bf Fig. 8}
Quartic fermion coupling,  $\kappa \vp^2 \bar{\psi} \psi$, and potential, 
$V=\lambda_4 \phi^4 /4$. The decrease of the amplitude is given by the envelope
function, $\left( 1 + 0.075 \alpha \sqrt{\lambda_4} \kappa^2 \vp_0^3 t 
\right)^{-1/3}$.
Fig.~8a: $m=0, \kappa =0.5, \lambda=1, \alpha=1.0$.
Fig.~8b: $m=0, \kappa =1.0, \lambda=1, \alpha=0.9$.

\newpage

\psfig{file=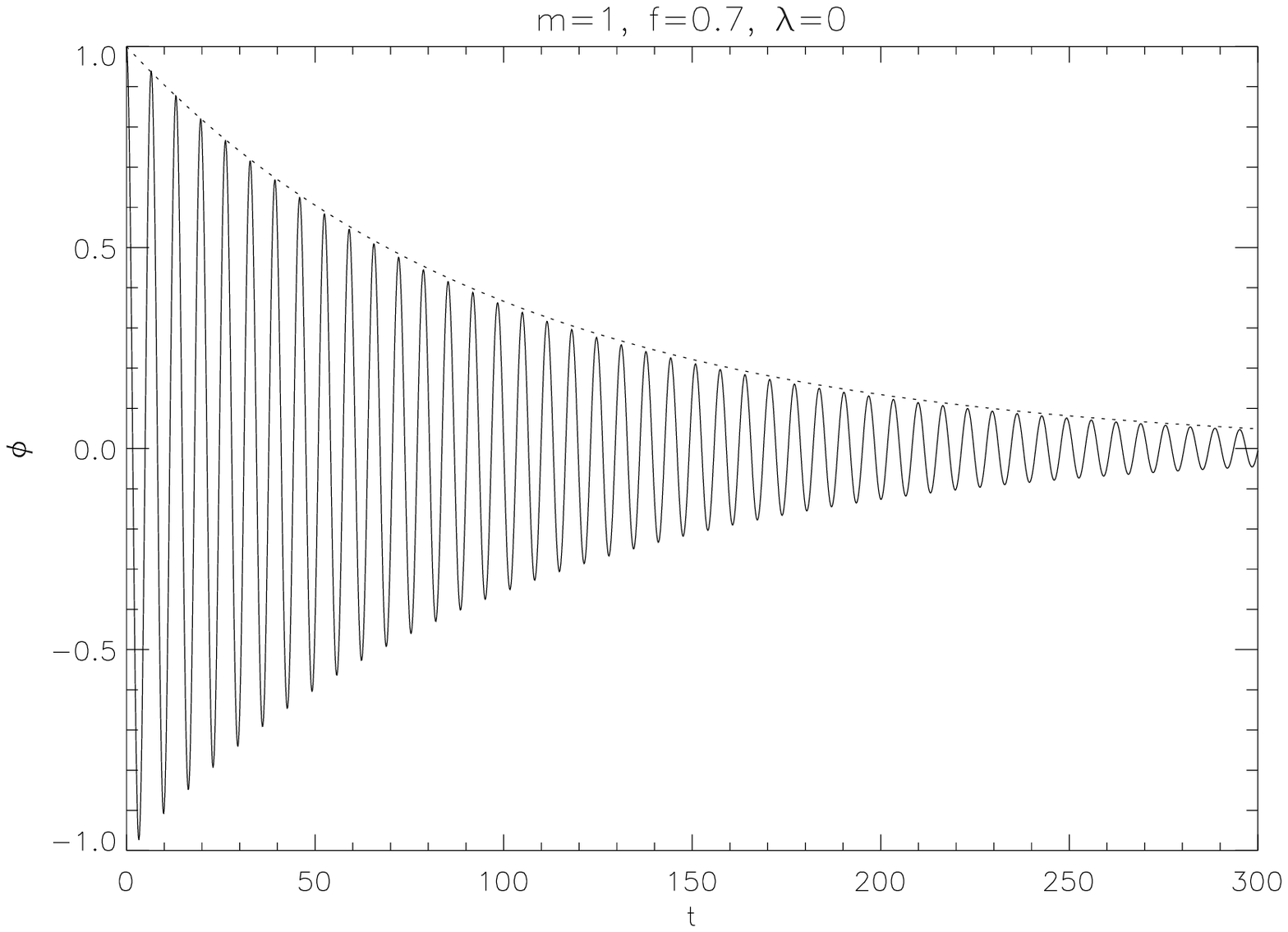,width=5in,height=3.5in}
\begin{center}
{\bf Figure 1a.}
\end{center}

\psfig{file=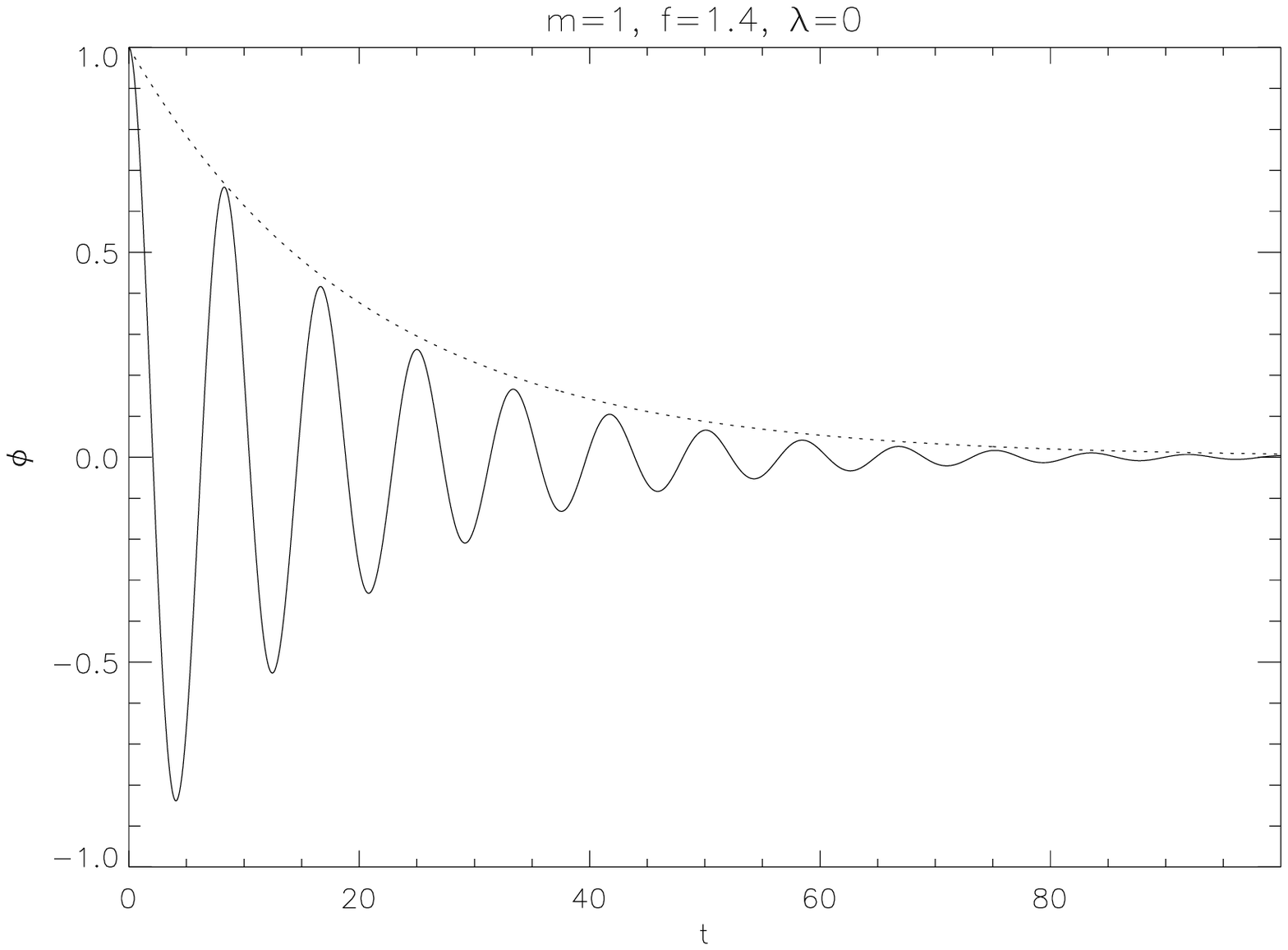,width=5in,height=3.5in}
\begin{center}
{\bf Figure 1b.}
\end{center}

\newpage

\psfig{file=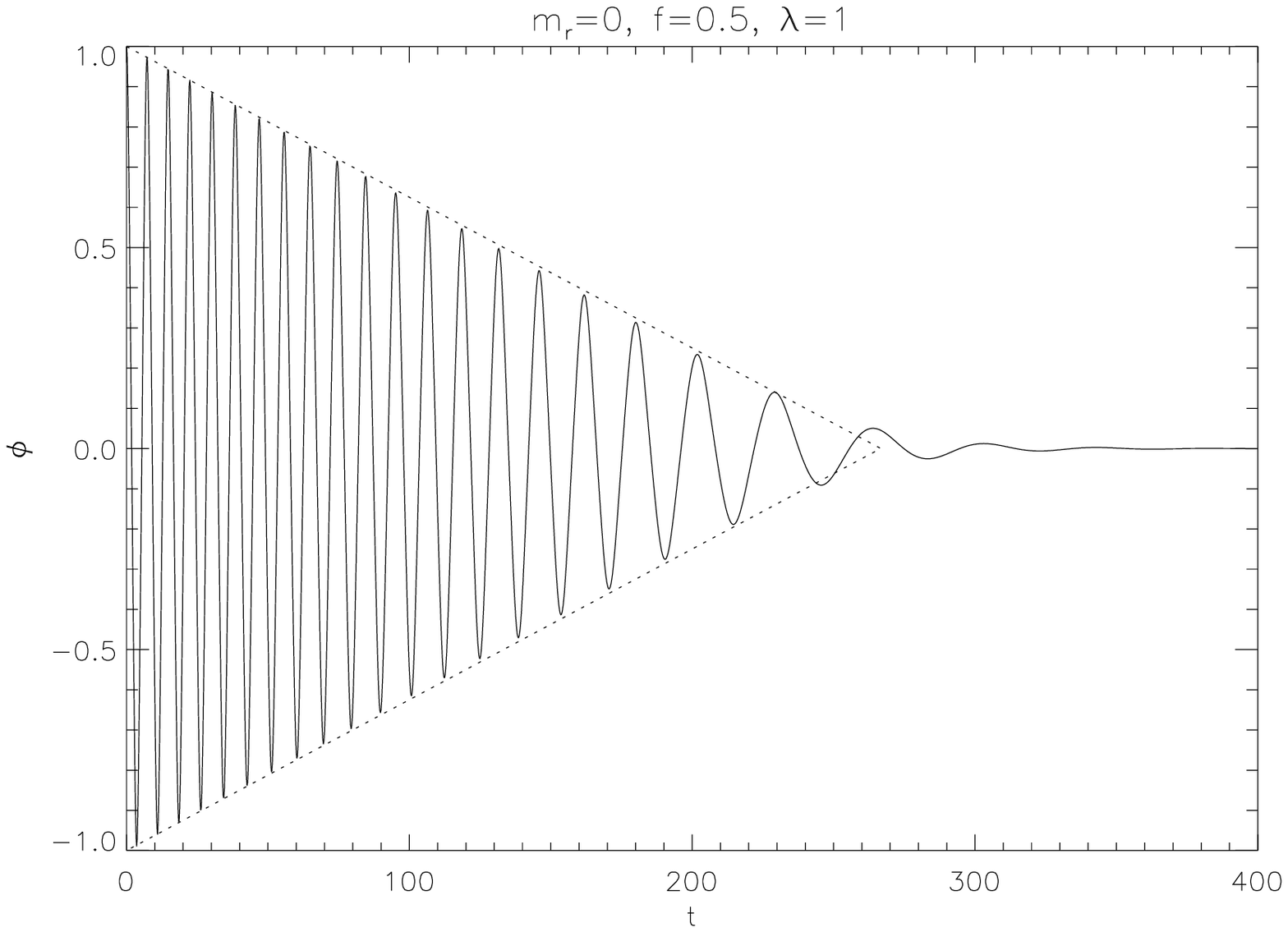,width=5in,height=3.5in}
\begin{center}
{\bf Figure 2a.}
\end{center}

\psfig{file=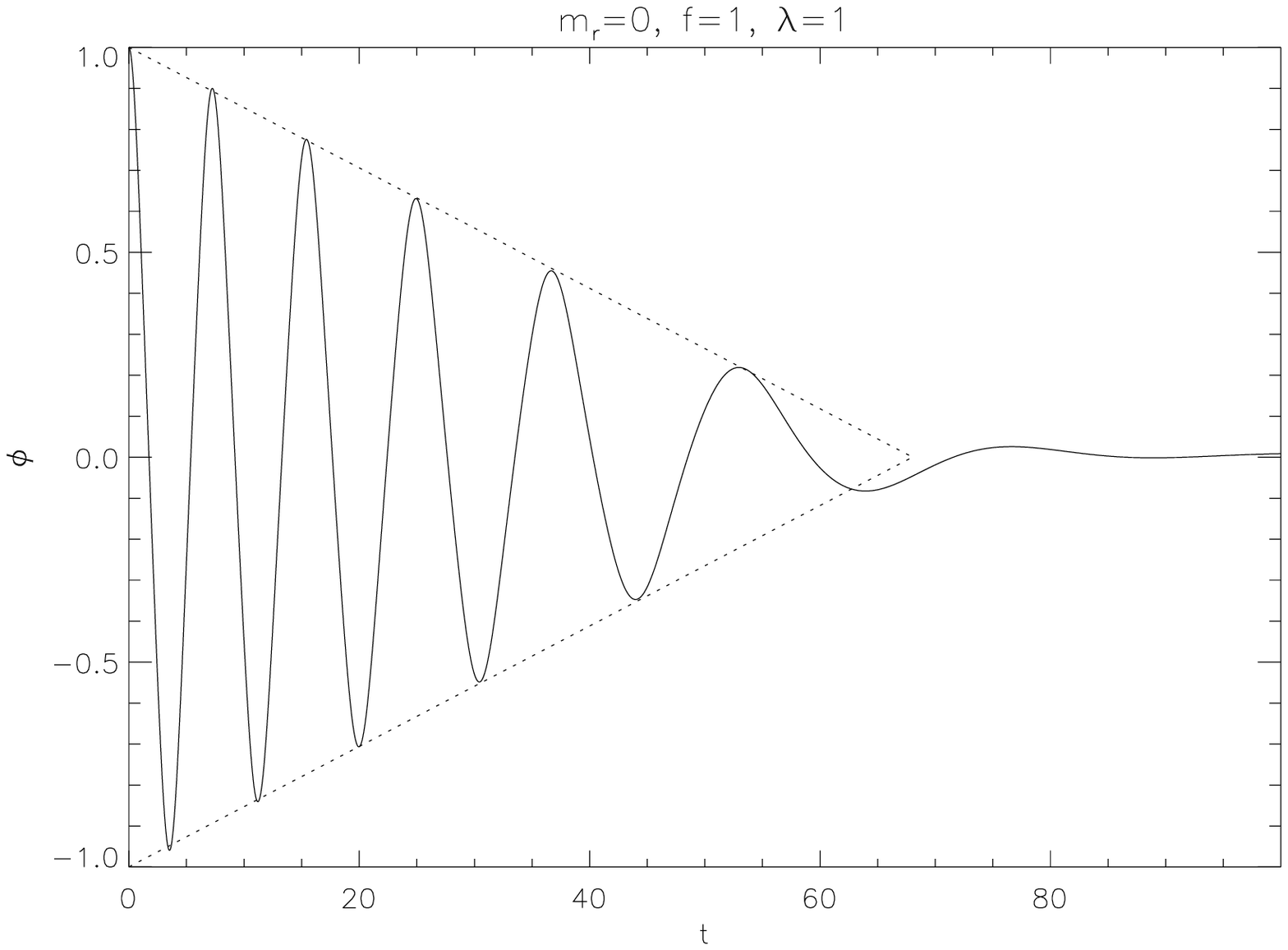,width=5in,height=3.5in}
\begin{center}
{\bf Figure 2b.}
\end{center}

\newpage

\psfig{file=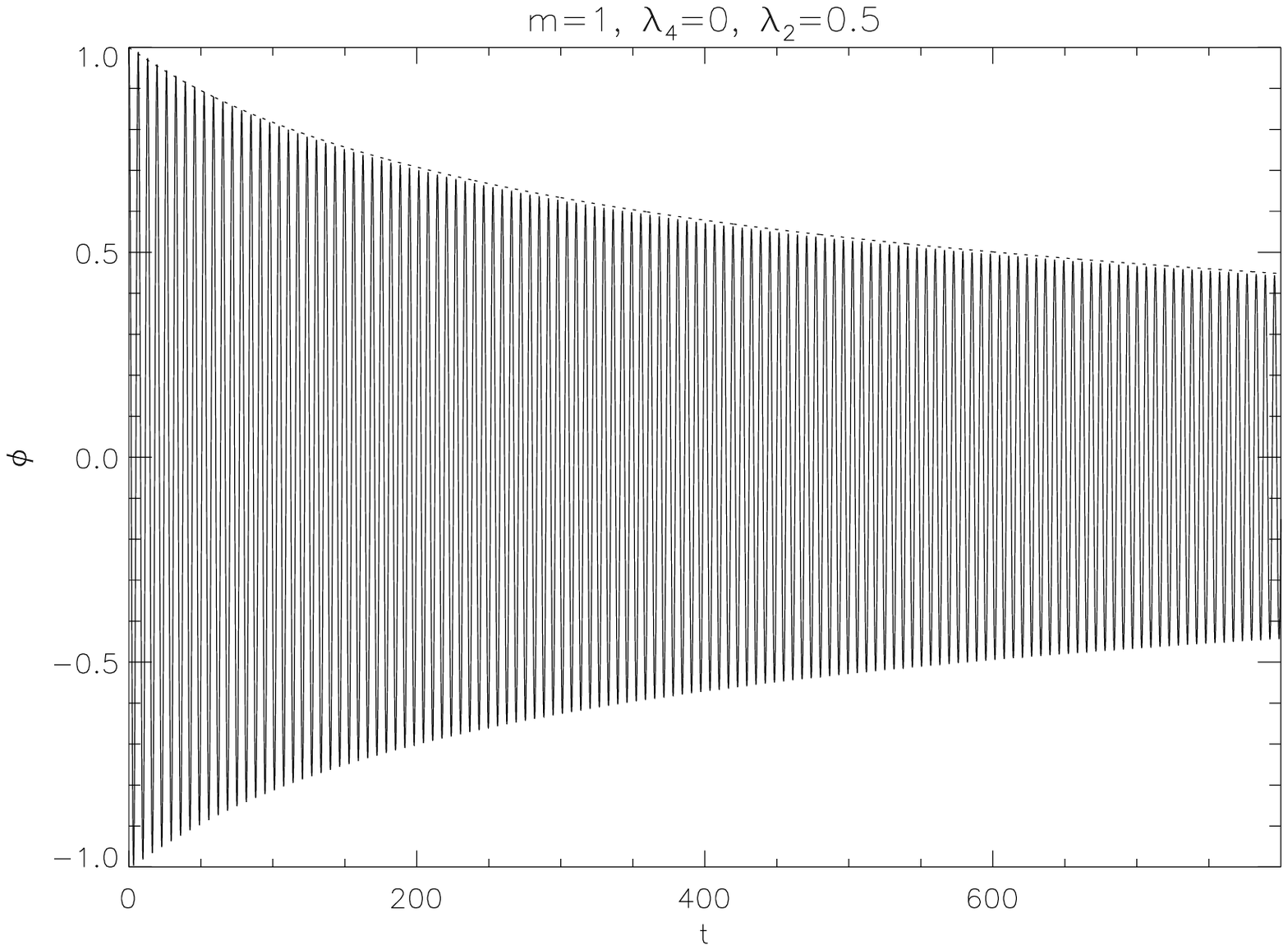,width=5in,height=3.5in}
\begin{center}
{\bf Figure 3a.}
\end{center}

\psfig{file=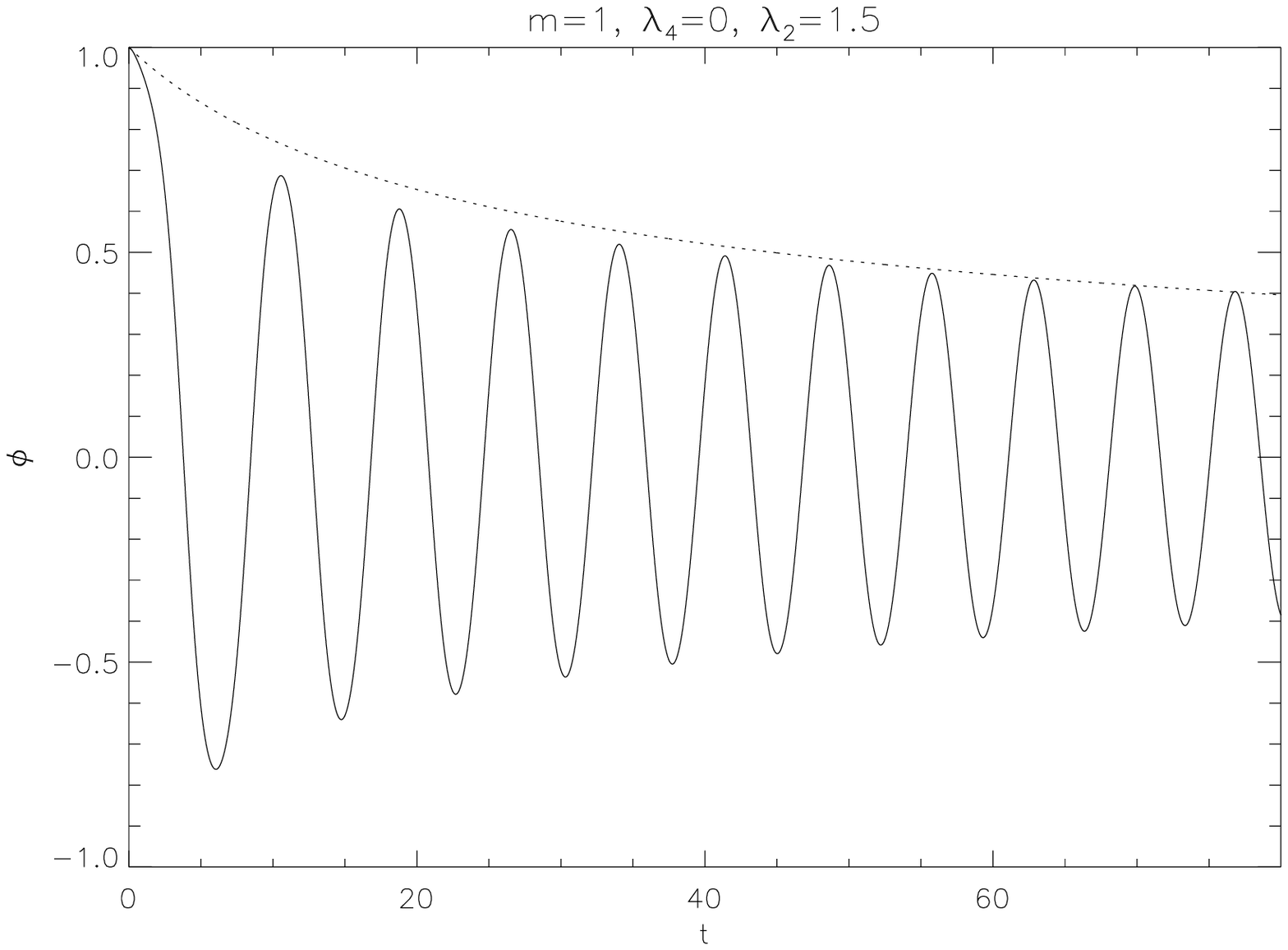,width=5in,height=3.5in}
\begin{center}
{\bf Figure 3b.}
\end{center}

\newpage

\psfig{file=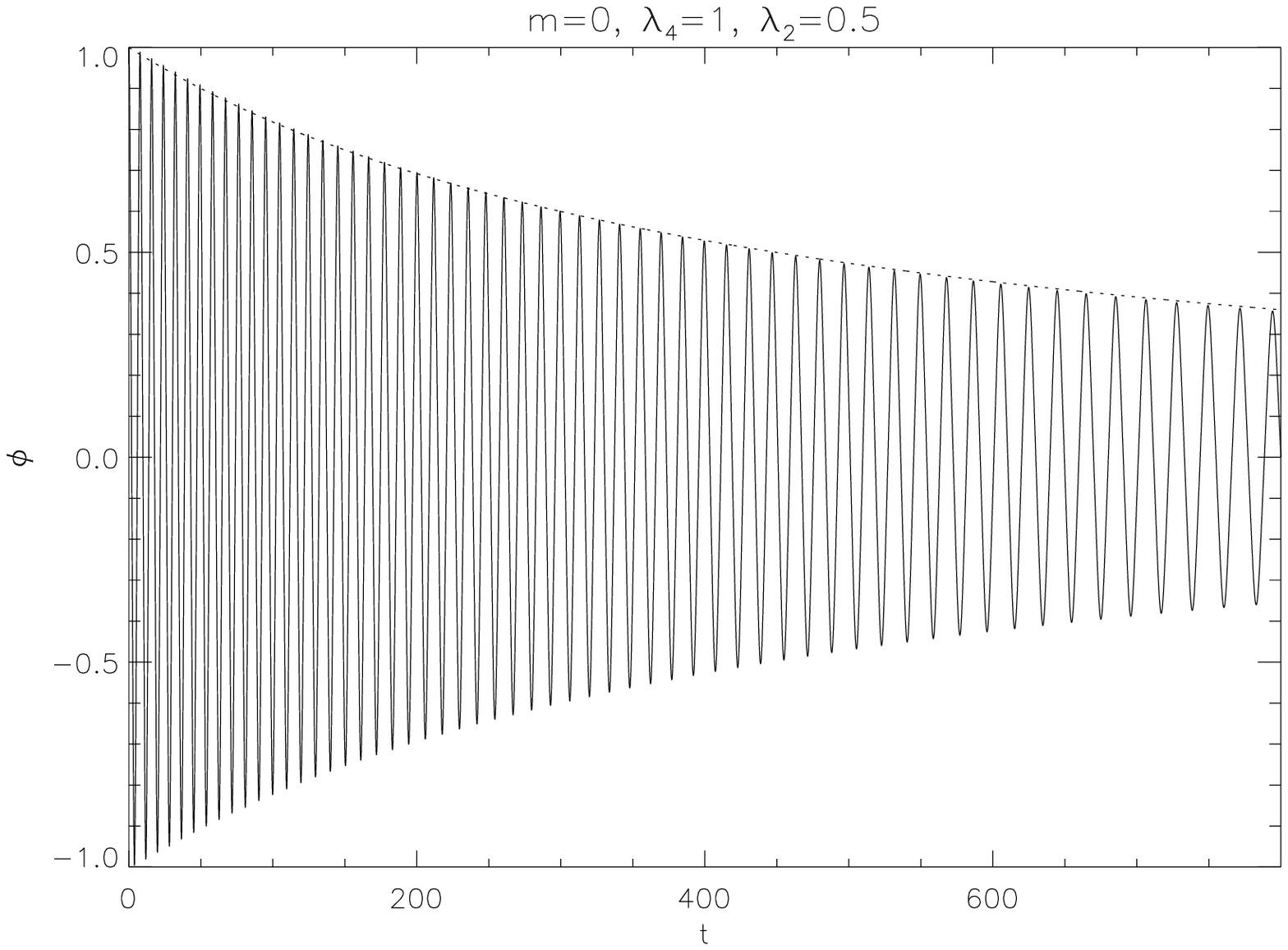,width=5in,height=3.5in}
\begin{center}
{\bf Figure 4a.}
\end{center}

\psfig{file=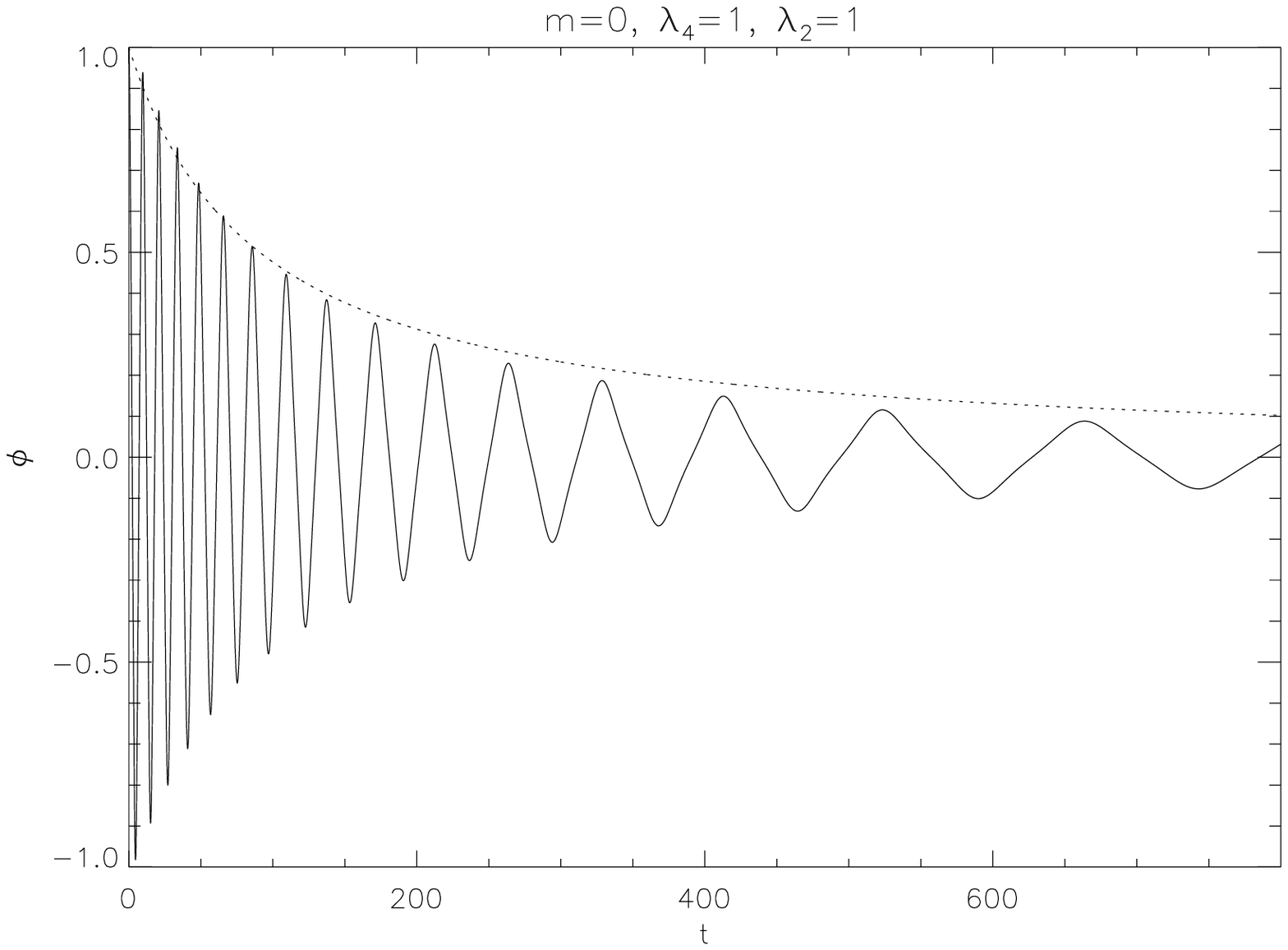,width=5in,height=3.5in}
\begin{center}
{\bf Figure 4b.}
\end{center}

\newpage

\psfig{file=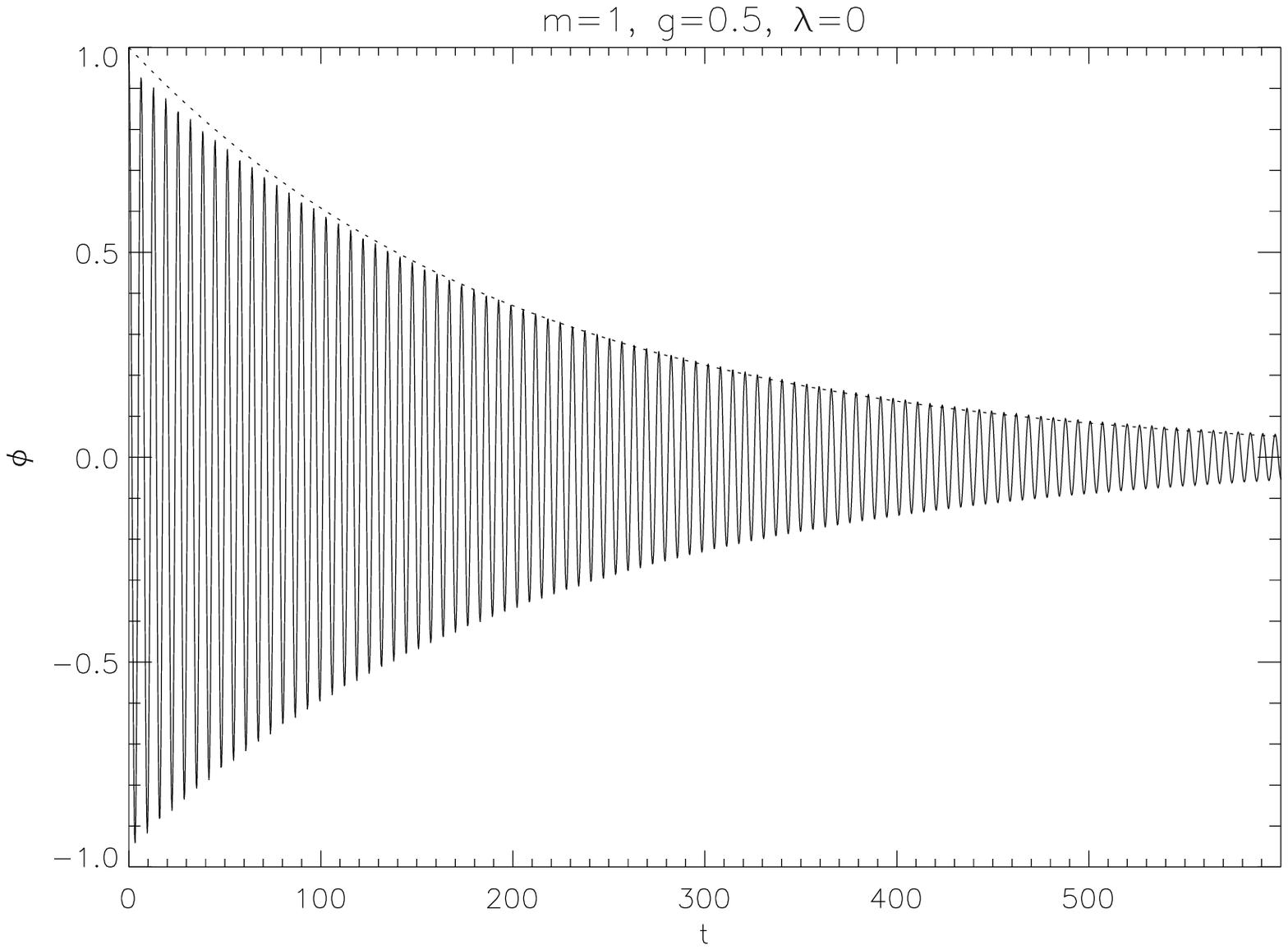,width=5in,height=3.5in}
\begin{center}
{\bf Figure 5a.}
\end{center}

\psfig{file=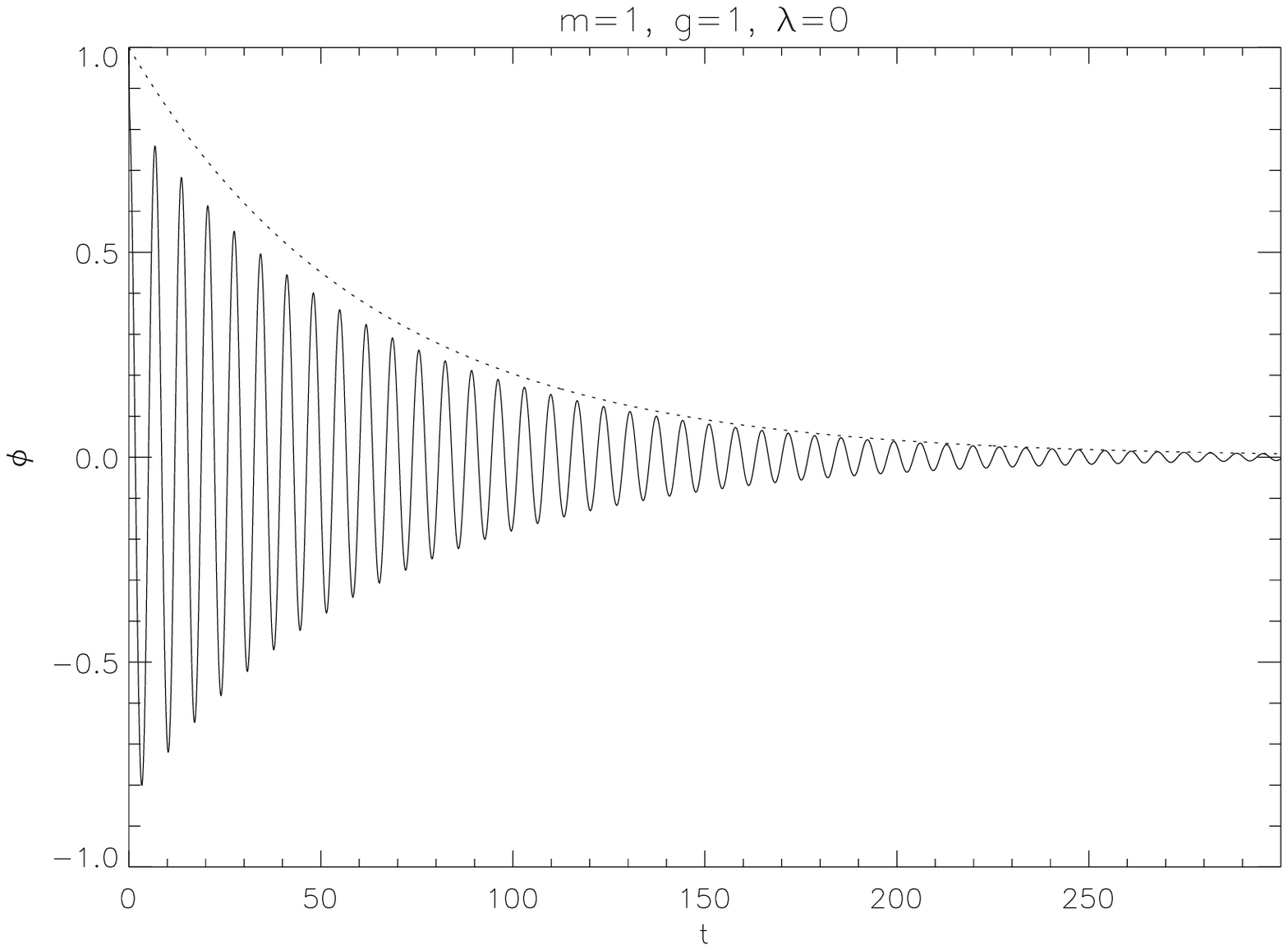,width=5in,height=3.5in}
\begin{center}
{\bf Figure 5b.}
\end{center}

\newpage

\psfig{file=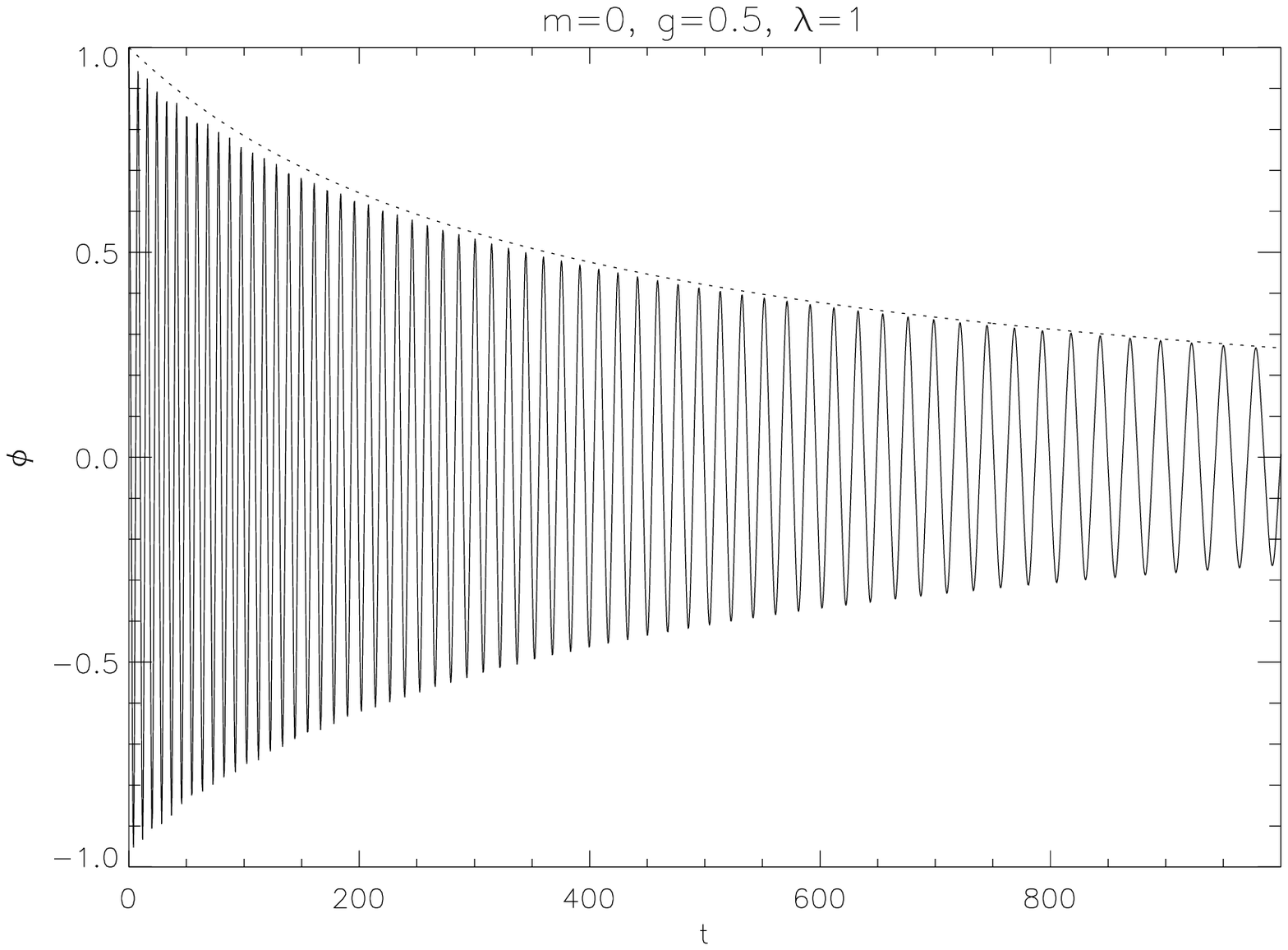,width=5in,height=3.5in}
\begin{center}
{\bf Figure 6a.}
\end{center}

\psfig{file=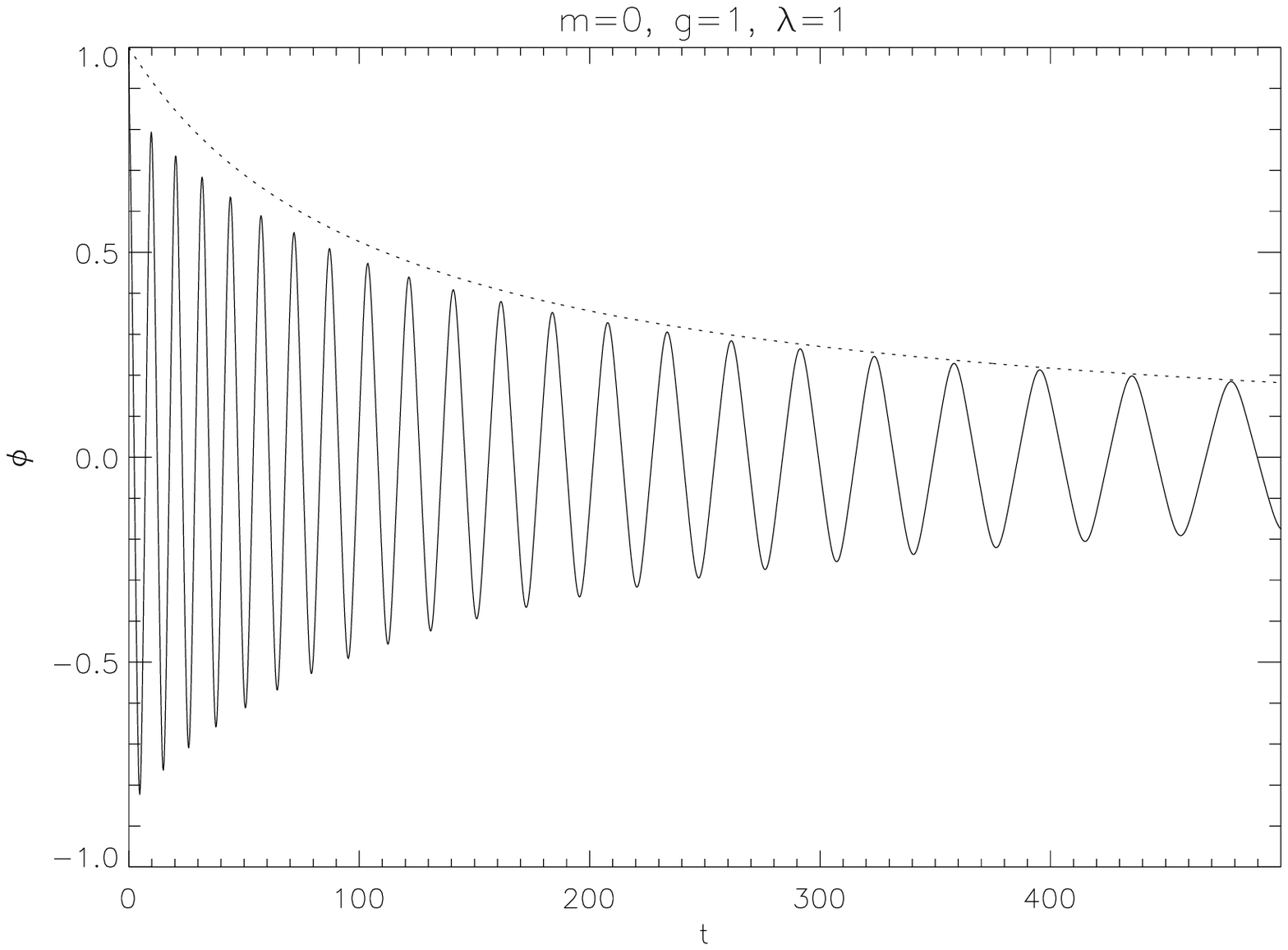,width=5in,height=3.5in}
\begin{center}
{\bf Figure 6b.}
\end{center}

\newpage

\psfig{file=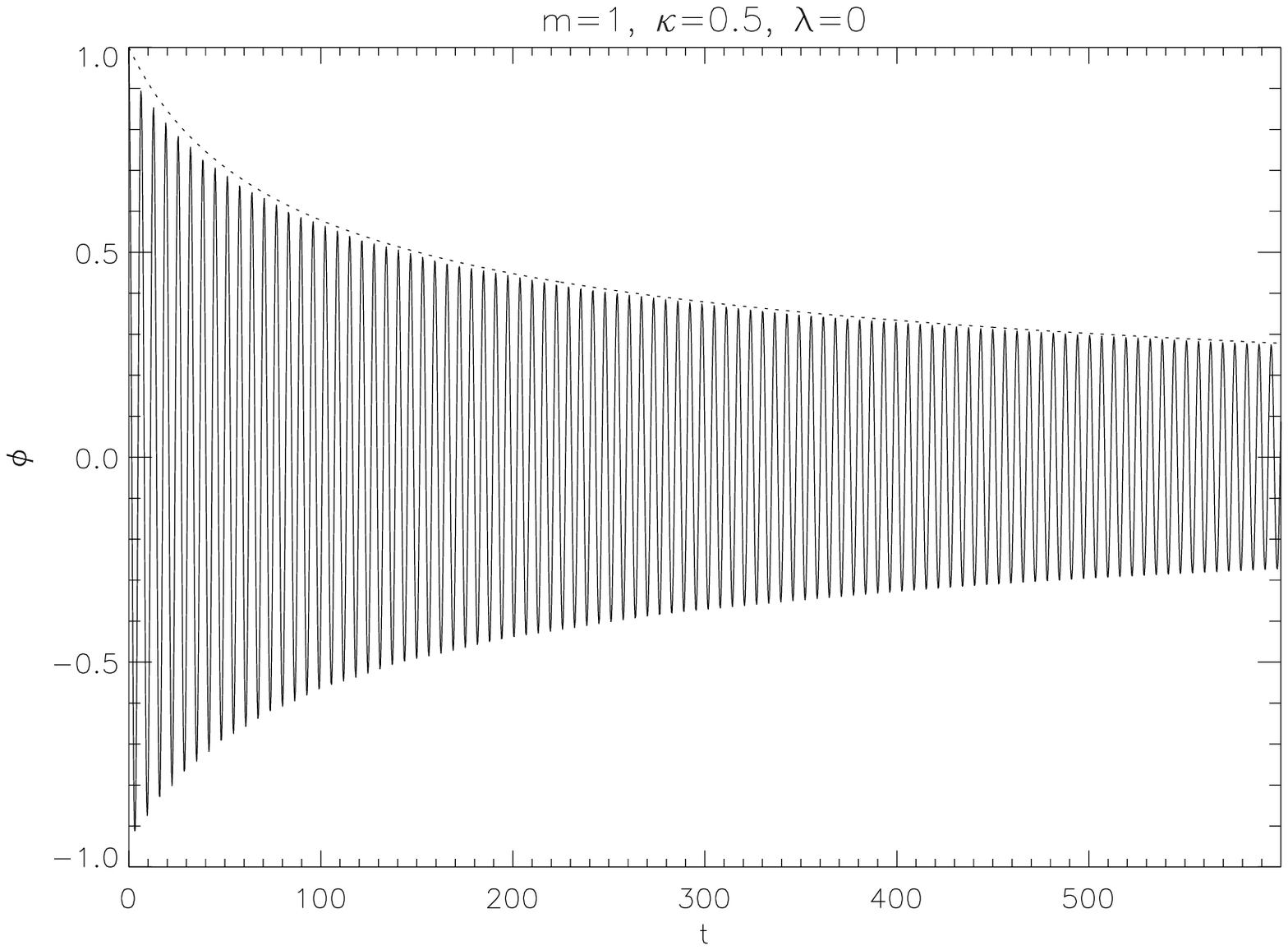,width=5in,height=3.5in}
\begin{center}
{\bf Figure 7a.}
\end{center}

\psfig{file=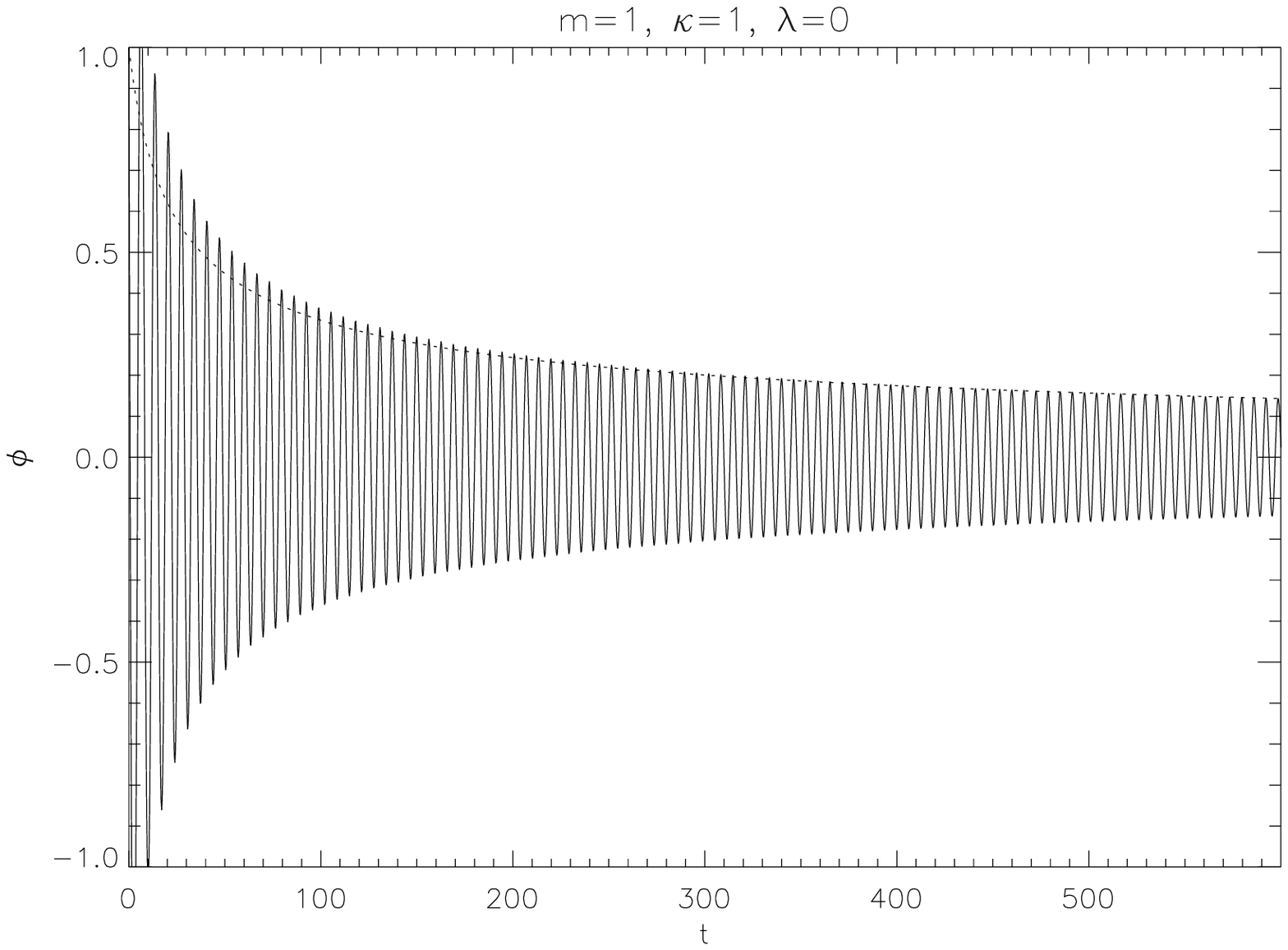,width=5in,height=3.5in}
\begin{center}
{\bf Figure 7b.}
\end{center}

\newpage

\psfig{file=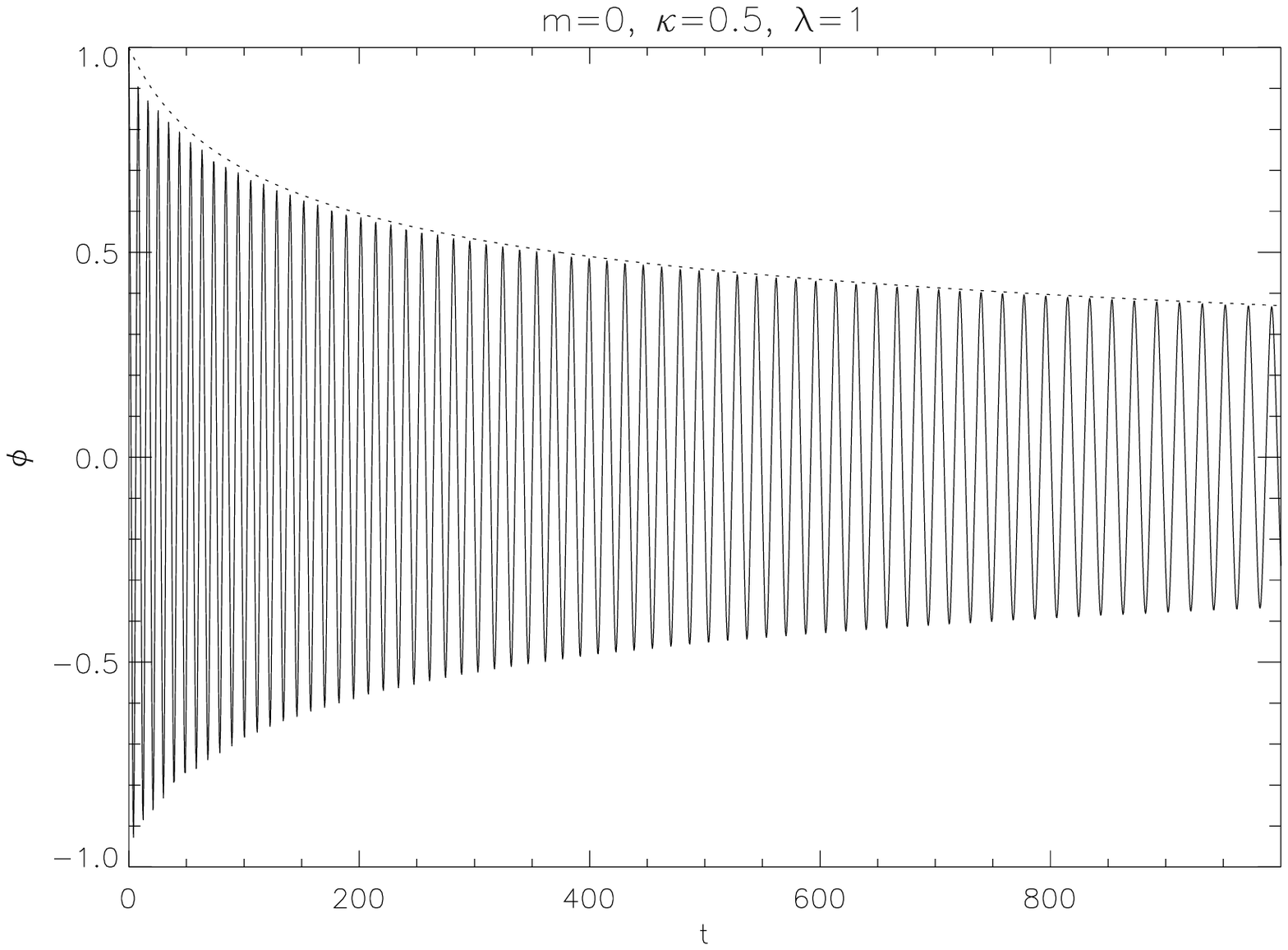,width=5in,height=3.5in}
\begin{center}
{\bf Figure 8a.}
\end{center}

\psfig{file=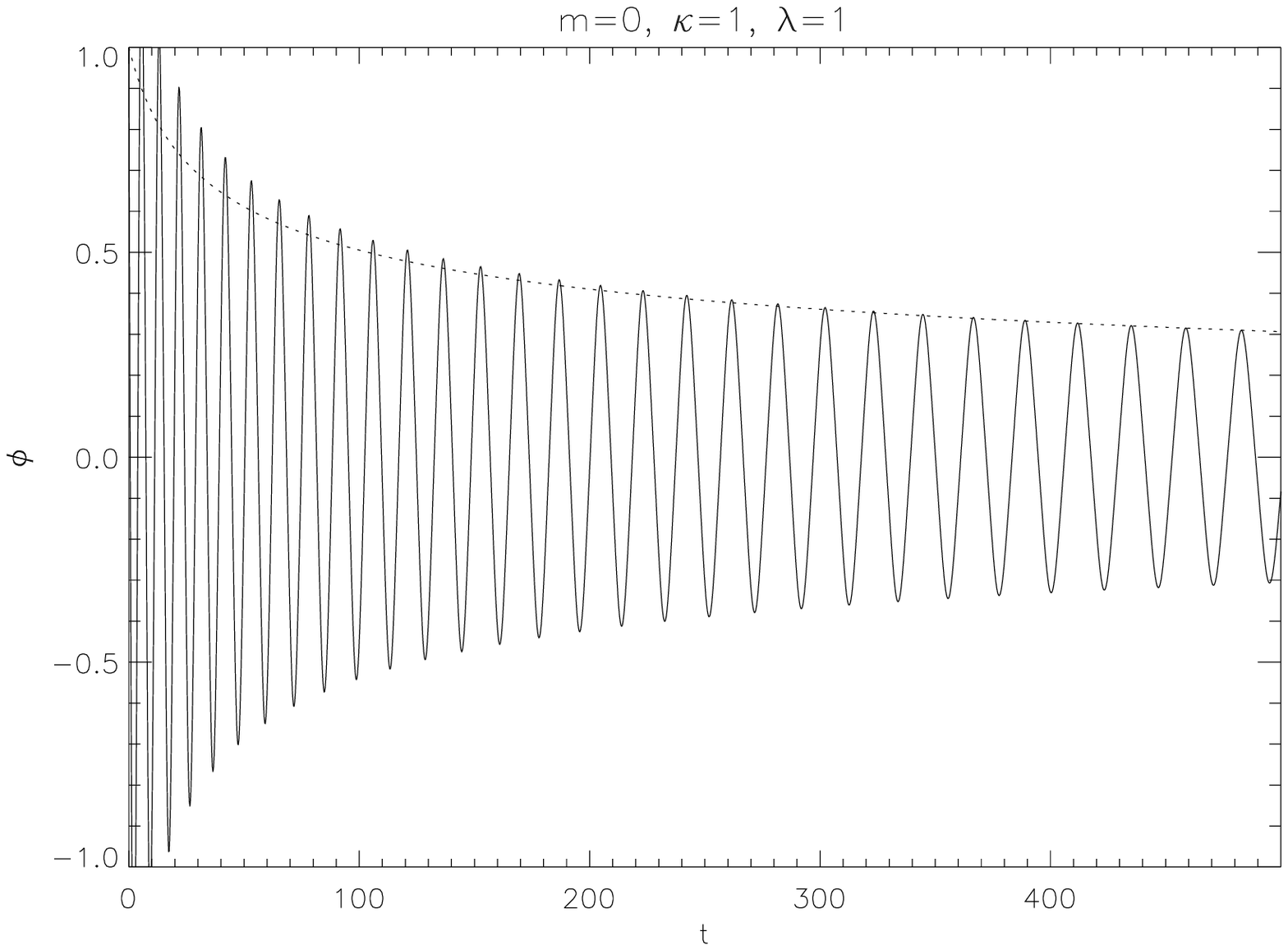,width=5in,height=3.5in}
\begin{center}
{\bf Figure 8b.}
\end{center}

\end{document}